\documentclass[twocolumn,preprintnumbers]{revtex4}

\usepackage{graphicx}% Include figure files
\usepackage{bm}
\usepackage{float}

\begin{document}

\title{Exact continuum theory of anti-Klein tunneling in bilayer graphene}

\author{P. A. Maksym$^{1,2}$}
\author{H. Aoki$^{1,3}$}

\affiliation{$^1$Department of Physics, University of Tokyo, Hongo, Tokyo
 113-0033, Japan\\
 $^2$School of Physics and Astronomy, University of Leicester, 
 Leicester LE1 7RH, UK\\
 $^3$Electronics and Photonics Research Institute,
 National Institute of Advanced Industrial Science and
  Technology (AIST), Tsukuba 305-8568, Japan}

\date{\today}

\begin{abstract}
  Exact conditions for anti-Klein transmission zeros are found analytically
  with a 4-component continuum approach which includes trigonal warping.
  Anti-Klein tunneling occurs at oblique incidence on steps and barriers
  with soft and hard walls as well as in the known case of normal incidence on
  a hard step. The necessary energy and angle of incidence depend on the
  crystallographic orientation of the step or barrier. At normal incidence
  on an armchair step in unbiased bilayer graphene, anti-Klein tunneling
  occurs because both the continuum and the tight binding Hamiltonians are
  invariant under layer and site interchange. At oblique incidence,
  anti-Klein tunneling is valley-dependent even in the absence of trigonal
  warping. An experimental arrangement that functions both as a detector of
  anti-Klein tunneling and a valley polarizer is suggested. There are
  cases where anti-Klein tunneling occurs in the 4-component theory but
  not in the 2-component approximation.
\end{abstract}

\maketitle
  
\section{Introduction}
Anti-Klein (AK) tunneling is the absence of tunneling at a potential step
in bilayer graphene (BLG). It was discovered theoretically
\cite{Katsnelson06} by using the 2-component approximation \cite{McCann13}
to the full 4-component continuum Hamiltonian and has been attributed to
the pseudospin of the 2-component states \cite{Katsnelson06,
  Park11, Gu11, Park12}. However the 4-component continuum Hamiltonian
cannot be expressed exactly in terms of a pseudospin vector. So can AK
tunneling occur in the 4-component continuum theory? We show that it can.
We also show that AK tunneling is valley asymmetric and may occur
at arbitrary potentials with both soft and hard walls. And we show further
that it occurs in a tight binding theory.

Absence of tunneling means that the transmission coefficient of a step or
barrier is \textit{exactly} zero. This happens at a p-n or n-p junction,
that is when the electron energy is in the conduction band on one side of a
potential interface and in the valence band on the other side. Within this
energy range, the transmission coefficient may vanish over an extended
range of energies \cite{Katsnelson06} or at a single critical energy
\cite{Park11}. Which case occurs depends on the structure and geometry of
the interface. We use 'AK tunneling' to mean zero transmission in these
cases and others we report here.

In the first work on AK tunneling \cite{Katsnelson06}, it was found that
zero transmission occurs at normal incidence on a potential step in
unbiased BLG. The transmission vanishes everywhere between the conduction
and valence band edges and the zero is exact within the 2-component
approximation without trigonal warping (TW). It occurs because pseudospin
conservation requires that the propagating plane wave incident on a step
matches onto an evanescent plane wave on the other side of the step.

Subsequently AK tunneling was found at normal incidence on a potential
step in biased BLG \cite{Park11}, again in the 2-component approximation
without TW. In this case the transmission vanishes at one critical energy
where the incident state matches onto an evanescent state on the other side
of the step. At this energy, the pseudospin conservation condition is that
the expectation values of the pseudospin of the incident and evanescent
states are identical.

The existence of \textit{exact} transmission zeros in the 2-component approximation
is a puzzle because the full 4-component Hamiltonian cannot be expressed in
terms of a pseudospin vector. To solve this puzzle, we find the condition
for AK tunneling in the 4-component continuum approach, including TW,
analytically. It turns out that AK tunneling at a potential step occurs
when a particular pair of evanescent wave polarization vectors on the left
and right sides of the step are orthogonal.

The orthogonality relation is a general condition for exact transmission
zeros. If it is evaluated with 4-component vectors, it gives the condition
for AK tunneling in the 4-component approach. If it is evaluated with
vectors found from the 2-component approximation it gives the condition for
AK tunneling in the 2-component approximation. Thus exact transmission
zeros can occur in both approaches but normally at different
\cite{footnote1} incidence conditions. This seems to solve the puzzle.

But what is the origin of the pseudospin conditions? In brief,
symmetry. When the step edge is parallel to an armchair direction (or
arbitrary direction with no TW), the 4-component continuum and tight
binding Hamiltonians for normal incidence are invariant under simultaneous interchange of layers
and sites. We call this swap symmetry and show that the swap quantum number
of the corresponding 4-component states is $\pm 1$, like the pseudospin.
The orthogonality relation and the swap symmetry lead to all the pseudospin
conditions found in the 2-component approximation. Thus we arrive at a
consistent and exact picture of AK tunneling that is valid in both the
4-component theory and the 2-component approximation.

However our objective is to go beyond this point and investigate the
physics of AK tunneling systematically. In the case of an arbitrary
potential, our orthogonality condition for AK tunneling at a hard
potential step generalizes to vanishing of the corresponding transfer
matrix element.  We use the orthogonality and transfer matrix conditions to
search for AK tunneling systematically. We find it not only in the well
known case of normal incidence on hard steps but also at oblique incidence
on steps and barriers with soft and hard walls. Further, because of TW, the
conditions for AK tunneling depend strongly on the crystallographic
orientation of the step or barrier. The occurrence of AK tunneling at
soft-walled potentials is particularly significant because these systems
are experimentally realizable.

Another very interesting feature of AK tunneling is that it is valley
asymmetric unless the transmission coefficient within each valley is
symmetric in the transverse momentum. The reason is that the polarization
vectors are valley-dependent and hence the critical transverse momentum
needed to satisfy the orthogonality relation is also valley-dependent. At
this critical momentum the valley asymmetry is large because the
transmission coefficient vanishes in only one of the valleys.
This effect may be used to make a valley polarizer.

The 4-component continuum theory is appropriate for our investigations
because experimentally realizable potentials vary slowly compared to the
length scale of the lattice. In addition, the continuum theory has the
advantage that it is easy to take account of the crystallographic
orientation of the step or barrier. The continuum and tight binding
Hamiltonians have identical swap symmetry so AK tunneling occurs in both
approaches. Valley mixing occurs only in the tight binding theory but is
quite weak. We have verified this in the case of normal
incidence on a hard armchair step in unbiased BLG. AK tunneling occurs as
in the continuum theory and the effect of valley mixing on the reflected
current is between $10^{-3}$ and $10^{-5}$ of the total current. For an
experimentally realistic soft step, the effect should be even smaller.

We derive the conditions for AK tunneling in Section
\ref{TheorySection}. We then present numerical results to show AK
tunneling occurs at arbitrary incidence on potential steps and barriers
(Section \ref{ExamplesSection}). In the same section we show that swap
symmetry results in AK tunneling at normal incidence on a step in unbiased
BLG and, in addition, detail the effects of bias, TW and crystallographic
orientation.  The valley dependence of AK tunneling is explained in
Section \ref{ValleySection} and in Section \ref{ExperimentalSection} we
suggest experimental arrangements for observing AK tunneling and for
generating valley polarized currents. The relation between the 4-component
theory and the 2-component approximation is explained in Section
\ref{2cSection} and our conclusions are summarized in Section
\ref{Discussion}. Appendix \ref{SymmetryAppendix} details transmission
coefficient relations that are used in Sections \ref{ExamplesSection} and
\ref{ValleySection}. Mathematical details of the relation between the
4-component theory and 2-component approximation are given in Appendices
\ref{UnbiasedAppendix} and \ref{BiasedAppendix}. The tight binding theory
of a hard armchair step is explained in Appendix \ref{TBAppendix}.

\section{Theory}
\label{TheorySection}
\subsection{Hamiltonian and plane wave states}
\label{4cHPWSection}
\begin{figure}
  \begin{center}
  \includegraphics[width=3.3cm]{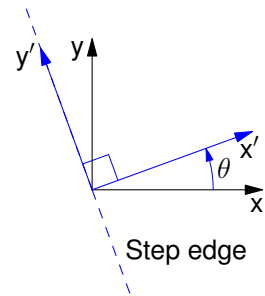}
  \caption{(Color online) Step edge (dashed blue line), crystallographic
    axes, $x,y$ (black arrows) and axes $x', y'$ fixed to step edge
    (blue arrows).}
\label{axisfig}
\end{center}
\end{figure}
We consider a step or barrier with edge normal at an angle $\theta$ to the
crystallographic $x$ axis. To find the transmission coefficient we use
co-ordinates $x',y'$ that are rotated by $\theta$ with respect to the
crystallographic co-ordinates, $x,y$ (Fig.~\ref{axisfig}).

The 4-component states are of form
$(\phi_{A1}, \phi_{B1}, \phi_{A2}, \phi_{B2})^T$ where the subscripts
denote sites within the BLG unit cell. The $K$-valley continuum
Hamiltonian, expressed in terms of $x',y'$, is
\begin{equation}
  H_{K} = 
    \left( \begin{array}{cccc}
    V_1 & v_0\pi_K^\dagger & -v_4\pi_K^\dagger  & v_3\pi_K e^{3i\theta}\\
    v_0\pi_K & V_1 + \Delta' & t & -v_4\pi_K^\dagger \\
    -v_4\pi_K  & t & V_2  + \Delta' & v_0\pi_K^\dagger\\
    v_3\pi_K^\dagger e^{-3i\theta} & -v_4\pi_K & v_0\pi_K & V_2 \\
    \end{array} \right), \label{hblg}
\end{equation}
where the unitary transformation
$\mathrm{diag}(e^{-i\theta}, 1, 1, e^{i\theta})$ has been used to reduce
the $\theta$ dependence to factors of the form $\exp(\pm 3i\theta)$
\cite{Maksym21}. Here $\pi_K = p_{x'} + ip_{y'}$,
$p_{x'}$ and $p_{y'}$ are momentum components and $v_0, v_3$ and $v_4$ are
velocities. $t$ is the interlayer coupling and $\Delta'$ is a small energy
shift of the interlayer coupled sites \cite{McCann13}. The step edge is
taken to be at $x'=0$. The potentials $V_i$ in layer $i$ become uniform far
away from the step or barrier edges. In $K'$, $\pi_K$ is
replaced by $\pi_{K'} \equiv -p_{x'} + ip_{y'}$ and $\theta$ by $-\theta$.

Plane wave states occur in the regions of uniform potential. In each valley
these states satisfy
\begin{equation}
  H \mathbf{e}_\alpha
  \exp(i\mbox{\boldmath$\kappa$}_\alpha \cdot \mathbf{r}) =
  E\mathbf{e}_\alpha\exp(i\mbox{\boldmath$\kappa$}_\alpha \cdot \mathbf{r}), \label{eigen}
\end{equation}
where $H$ is the appropriate valley Hamiltonian, $E$ is the energy,
$\mathbf{e}_\alpha$ is a polarization vector, $\alpha$ is a mode index,
$\mathbf{r} = (x',y')$ and
$\mbox{\boldmath$\kappa$}_\alpha = (k_\alpha, k_{y'})$ is the
$\mathbf{k}$-vector and $k_\alpha$ is its $x'$ component. The plane waves
may be propagating or evanescent.

To find $k_\alpha$ and the polarization vectors as a function of $E$ and
$k_{y'}$ we re-write Eq.~(\ref{eigen}) as an eigenvalue equation for
$p_\alpha \equiv \hbar k_\alpha$. This gives
\begin{equation}
  v_{x'}^{-1}(W + p_{y'} v_{y'})\mathbf{e}_\alpha
  = -p_\alpha \mathbf{e}_\alpha, \label{eigkx}
\end{equation}
where $p_{y'} = \hbar k_{y'}$,
\begin{equation}
  W = 
    \left( \begin{array}{cccc}
    V_1 - E& 0& 0 & 0\\
    0& V_1 + \Delta' - E& t & 0\\
    0 & t & V_2 + \Delta' - E & 0\\
    0 & 0 & 0& V_2 - E \\
    \end{array} \right) \label{Vdef}
\end{equation}
and the velocity operators in the $K$-valley are
\begin{equation}
  v_{x'} = 
    \left( \begin{array}{cccc}
    0 & v_0& -v_4  & v_3 e^{3i\theta}\\
    v_0& 0 & 0 & -v_4 \\
    -v_4  & 0 & 0 & v_0\\
    v_3 e^{-3i\theta} & -v_4 & v_0& 0 \\
    \end{array} \right) \nonumber \\
    \label{vxdef}
\end{equation}
and
\begin{equation}
  v_{y'} = 
    \left( \begin{array}{cccc}
    0 & -iv_0& iv_4  & iv_3 e^{3i\theta}\\
    iv_0& 0 & 0 & iv_4 \\
    -iv_4  & 0 & 0 & -iv_0\\
    -iv_3 e^{-3i\theta} & -iv_4 & iv_0& 0 \\
    \end{array} \right).
    \label{vydef}
\end{equation}
In the $K'$ valley $\theta$ is replaced by $-\theta$ and the sign of the
velocity parameters changes in $v_{x'}$.

The matrix on the left hand side of Eq.~(\ref{eigkx}) is a general complex
matrix hence its left eigenvectors, $\mathbf{f}^\dagger_\alpha$, and right
eigenvectors, $\mathbf{e}_\alpha$, form a biorthogonal set, that is
\begin{equation}
  \mathbf{f}^\dagger_\alpha \cdot \mathbf{e}_\beta = \delta_{\alpha \beta},
  \label{biorthodef}
\end{equation}
where the $\mathbf{e}$ vectors are normalized so that
$\mathbf{e}^\dagger_\alpha \cdot \mathbf{e}_\alpha = 1$.

The biorthogonality relation, Eq.~(\ref{biorthodef}), is valid for any general
complex matrix but in the special case of the matrix in Eq.~(\ref{eigkx}),
there is also a relation between the $\mathbf{e}$ vectors and the
$\mathbf{f}^\dagger$ vectors. By taking the Hermitean conjugate of
Eq.~(\ref{eigkx}) it can be shown that
\begin{equation}
\mathbf{f}^\dagger(k_\alpha) =
N_{k_\alpha}\mathbf{e}^\dagger(k^*_\alpha) v_{x'},
\label{ferel}
\end{equation}
 where $N_{k_\alpha}$
is a normalization constant and the $k_\alpha$ are either real or form
complex conjugate pairs. Then it follows from  Eq.~(\ref{biorthodef}) that
\begin{equation}
  \mathbf{e}^\dagger(k_\alpha) v_{x'} \mathbf{e}(k_\beta) \propto
  \delta_{k^*_\alpha k_\beta}.
  \label{orthojdef}
\end{equation}
That is, the $\mathbf{e}$ vectors are orthogonal with respect to the $x'$
component of the velocity and hence the $x'$ component of the current.

The physical consequence of this orthogonality is that in a superposition
of plane wave states there is no interference between the currents carried
by the propagating states and if a tunneling current is present it is
spatially uniform. Orthogonality relations similar to Eq.~(\ref{orthojdef})
have been found in a $\mathbf{k}\cdot\mathbf{p}$ theory of semiconductor
superlattices \cite{Smith86} and a tight binding theory of potential
barriers in graphene \cite{Chen16}. In an earlier paper \cite{Maksym21},
we used Eq.~(\ref{orthojdef}) to
simulate scattering in BLG numerically but without
presenting the proof given here.

\subsection{AK tunneling at hard steps}
\label{4cTRSection}

The transmission and reflection coefficients can be found easily by using
biorthogonality. We explain this first for the case when AK tunneling may
occur, i.e. when there are two propagating modes and two evanescent modes
on both sides of the step.

A plane wave is taken to be incident from the left of the step. The wave
functions $\psi_l$ and $\psi_r$ on the left and right sides of the step are
\begin{eqnarray}
  \psi_l &=& [\mathbf{e}_{1l}e^{ik_{1l} x'} +
    r_2\mathbf{e}_{2l}e^{ik_{2l} x'} + r_4\mathbf{e}_{4l}e^{ik_{4l} x'}]
  e^{ik_{y'} y'}\hspace{-1mm},\hspace{3mm}\label{psileft}\\
  \psi_r &=& [t_1 \mathbf{e}_{1r}e^{ik_{1r} x'} +
    t_3\mathbf{e}_{3r}e^{ik_{3r} x'}]e^{ik_{y'} y'}\hspace{-1mm},
  \label{psiright}
\end{eqnarray}
where the $t_i$ are transmitted amplitudes and $r_i$ are reflected
amplitudes. Mode 1 is right propagating, mode 2 is left propagating, mode
3 is right decaying, mode 4 is left decaying and the subscripts $l$ and $r$
denote the left and right sides of the step. The wave function must be
continuous at the step edge. Hence
\begin{equation}
  \mathbf{e}_{1l} + r_2\mathbf{e}_{2l} + r_4\mathbf{e}_{4l} =
  t_1 \mathbf{e}_{1r} + t_3\mathbf{e}_{3r}.
  \label{psimatch}
\end{equation}
Equations for $t_1$ and $t_3$ are obtained by applying the biorthogonality
condition to Eq.~(\ref{psimatch}). Thus
\begin{eqnarray}
  \mathbf{f}^\dagger_{1l}\cdot\mathbf{e}_{1r} t_1 +
  \mathbf{f}^\dagger_{1l}\cdot\mathbf{e}_{3r} t_3 &=& 1 \label{teq1}\\
  \mathbf{f}^\dagger_{3l}\cdot\mathbf{e}_{1r} t_1 +
  \mathbf{f}^\dagger_{3l}\cdot\mathbf{e}_{3r} t_3 &=& 0. \label{teq2}
\end{eqnarray}
The coefficient matrix in these equations must be non-singular and this
excludes the possibility that $\mathbf{f}^\dagger_{3l}\cdot\mathbf{e}_{1r} = 0$
when $\mathbf{f}^\dagger_{3l}\cdot\mathbf{e}_{3r} = 0$. Hence when
\begin{equation}
  \mathbf{f}^\dagger_{3l}\cdot\mathbf{e}_{3r} = 0,\label{antikcon}
\end{equation}
the transmission coefficient, $t_1$, vanishes. Eq.~(\ref{antikcon}) is
the orthogonality condition mentioned in the introduction and
is the exact condition for AK tunneling at a hard potential step.
It may be satisfied because of swap symmetry or for critical
values of the incidence parameters (Section \ref{ExamplesSection}).

The reflection coefficients may also be obtained from Eq.~(\ref{psimatch})
and are given by
\begin{eqnarray}
  r_2 &=& \mathbf{f}^\dagger_{2l}\cdot\mathbf{e}_{1r} t_1 +
  \mathbf{f}^\dagger_{2l}\cdot\mathbf{e}_{3r} t_3 \label{req1}\\
  r_4 &=& \mathbf{f}^\dagger_{4l}\cdot\mathbf{e}_{1r} t_1 +
  \mathbf{f}^\dagger_{4l}\cdot\mathbf{e}_{3r} t_3. \label{req2}
\end{eqnarray}

In deriving Eqs.~(\ref{teq1}), (\ref{teq2}), (\ref{req1}) and (\ref{req2}),
we have focused on the case of two propagating modes and two evanescent
modes however the polarization vectors are biorthogonal in \textit{all}
cases and the number of modes does not change.
Hence Eqs.~(\ref{teq1}), (\ref{teq2}), (\ref{req1}), (\ref{req2}) are
always valid; the only case dependence is in the meaning of the mode indices.
%in all cases, provided the mode indices are defined appropriately,
Thus biorthogonality provides an easy way of finding the transmission
and reflection coefficients but as far as we know this has not been
reported before.

\subsection{AK tunneling at soft steps and arbitrary potential barriers}
\label{SoftSection}

The condition for AK tunneling at a hard step,
Eq.~(\ref{antikcon}), can be generalized to soft steps and arbitrary
potential barriers by using a transfer matrix \cite{Maksym21} to find the
transmission coefficients. The transfer matrix $M$ relates the amplitudes
of the waves on the left and right sides of the system,
$D(x_l)\mathbf{a}_l = M D(x_r)\mathbf{a}_r$, where $\mathbf{a}_l =
(\mathbf{r}^T, \mathbf{i}^T)^T$, $\mathbf{a}_r = (\mathbf{x}^T,
\mathbf{t}^T)^T$. Here $\mathbf{i}$ is a vector of incident wave
amplitudes, $\mathbf{r}$ is a vector of reflected wave amplitudes,
$\mathbf{t}$ is a vector of transmitted wave amplitudes, $\mathbf{x}$ is a
vector of the amplitudes of waves incident from the right and $D(x')$ is a
diagonal matrix of phase factors, $\exp(ik_i x')$.

The transmission coefficients satisfy equations analogous to Eqs.~(\ref{teq1})
and (\ref{teq2}),
\begin{eqnarray}
  M_{11} t_1 e^{i k_{1r} x'_r} + M_{13} t_3 e^{i k_{3r} x'_r} &=&
  e^{i k_{1l} x'_l} \label{mteq1}\\
  M_{31} t_1 e^{i k_{1r} x'_r} + M_{33} t_3 e^{i k_{3r} x'_r} &=& 0.
  \label{mteq2}
\end{eqnarray}
When
\begin{equation}
  M_{33} = 0 \label{mantikcon},
\end{equation}
the transmission coefficient, $t_1$, vanishes. Eq.~(\ref{mantikcon}) is the
transfer matrix condition mentioned in the introduction and is the exact
condition for AK tunneling at an arbitrary potential step or
barrier. Eq.~(\ref{mantikcon}) shows that AK tunneling may occur
but numerical calculations of $M_{33}$ are needed to check whether it does
occur. This is a difficult computational problem as large numerical errors
accumulate because of the growing exponential contributions to the transfer
matrix. This can be avoided by computing the transmission coefficient and
locating its zeros instead of searching for the zeros of $M_{33}$.

However $M_{33}$ can be computed accurately in the exceptional case of a thin
barrier which consists of a spatially uniform potential with hard edges.
In this case the transfer matrix elements are
\begin{equation}
  M_{\alpha\beta} = \mathbf{f}^\dagger_{\alpha l} \cdot \left[
  \sum_j \mathbf{e}_{jc} \exp(-ik_{jc} w) \mathbf{f}^\dagger_{jc}
  \right] \cdot \mathbf{e}_{\beta r},
  \label{mopdef}
\end{equation}
where $w$ is the barrier width and the subscript $c$ denotes polarization
vectors in the center of the barrier. The mathematical form of
Eq.~(\ref{mopdef}) is a consequence of biorthogonality. This form is valid
for arbitrary barrier widths but can be used to compute the transfer matrix
elements accurately only when the width is small.

\section{Examples of AK tunneling}
\label{ExamplesSection}
\begin{figure}
  \begin{center}
    \hspace{-4mm}
  \includegraphics[width=4.1cm, angle=0]{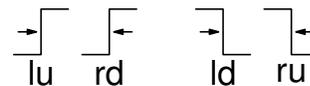}
  \caption{The four step configurations in each valley (schematic). For
    clarity, the bias potential is not shown. The
    arrows indicate the direction of incidence.}
\label{stepconfigsfig}
\end{center}
\end{figure}

In this section we give examples of AK tunneling in the 4-component,
continuum theory. Steps in unbiased BLG are discussed in Section
\ref{UnbiasedSection}, steps in biased BLG in \ref{BiasedSection} and
barriers in \ref{BarrierSection}. We also explain why AK tunneling at
normal incidence in unbiased BLG results from the swap symmetry of the
Hamiltonian (\ref{SymmetrySection}).

Transmission coefficients in BLG have novel features that result from
strong TW. When the constant energy contours are warped, the gradient of
$E(\mathbf{k})$ is no longer parallel to $\mathbf{k}$ so the current
carried by a Bloch state is also not parallel to $\mathbf{k}$. Further,
when there are points of inflection on the contour, several Bloch states
with distinct $\mathbf{k}$-vectors may contribute to the total current in a
particular direction \cite{Maksym21}. Thus multiple incident states may
occur and even when there is only one incident state it may couple to two
distinct propagating states on the exit side of a step. A similar situation
may occur without TW in biased BLG because of its Mexican hat band
structure.

When there is one incident state, the transmission coefficient is
\begin{equation}
  T = \frac{1}{j_{x1l}}\left(|t_1|^2 j_{x1r} + |t_3|^2 j_{x3r}\right),
  \label{tcoeffdef}
\end{equation}
where $j_{x1l}$ is the current carried by the incident state and $j_{x1r}$
and $j_{x3r}$ are transmitted state currents. When there is only one
propagating state on the exit side, $j_{x3r}$ vanishes because mode 3 is
then evanescent but when there are two propagating states $j_{x3r}$ is not
zero. Thus Eq.~(\ref{tcoeffdef}) gives the transmission coefficient in both
cases. In this work, we have found the case of two propagating transmitted
states only in Fig.~\ref{oblincfig} (left) and only in a very small range of
incidence angles (see figure caption). We have not found the case of
several incident states although this case can occur \cite{Maksym21} and is
relevant to experiment. It is discussed further in Section
\ref{ExperimentalSection}.

There are 4 step configurations in each valley, because carriers may be
incident from the left or right and may encounter an up step or a down step
(Fig.~\ref{stepconfigsfig}). This gives 8 possible transmission
coefficients when multiple propagating states do not occur and more
otherwise. However these transmission coefficients are related by symmetry
and all of them have similar features. We detail only the case of the
\textsf{lu} configuration in the $K$ valley. The relations between the
transmission coefficients are explained in Appendix \ref{SymmetryAppendix}.

$T$ is a function of $E$ and one variable related to the angle of
incidence. This variable can be either $k_{y'}$, or the polar angle of the
incident state $\mathbf{k}$-vector, $\phi_k$ (k-incidence angle) or the
polar angle of the incident current, $\phi_c$ (current incidence
angle). These angles are different in the presence of TW because the
current is not parallel to $\mathbf{k}$. We plot $T$ as a function of $E$,
$\phi_k$ or $k_{y'}$. However $\phi_c$ is relevant to experiments in the
ballistic transport regime so in the figure captions we give the values of
$\phi_c$ and $\phi_k$ at which AK zeros occur.

To find the AK condition we normally use bisection to locate the zeros of
$\mathbf{f}^\dagger_{3l}\cdot\mathbf{e}_{3r}$ or $M_{33}$. This method
brackets the roots of a function so we can be sure that a root exists
between the brackets that it returns. We stop bisecting when these brackets
differ by a number close to 64-bit precision. In the case of thin barriers,
$w \alt 150$ nm, with hard walls, we use Eq.~(\ref{mopdef}) to find $M_{33}$.
For thicker barriers or systems with soft walls, we use an $S$-matrix
method \cite{Maksym21} to search for minima of $T$. The minimum
value found in all cases is $<10^{-9}$.

Throughout this work we use '$\sim$' and '$=$' to distinguish incidence
parameters that are found numerically from incidence parameters that are
input to our codes. '$\sim$' followed by a number with 4 significant digits
indicates a parameter found numerically while '$=$' followed by a number
gives an exact input value.

The Hamiltonian parameters in meV \cite{Maksym21, McCann13} are: $\gamma_0
= 3160$, $\gamma_3 = 380$, $\gamma_4 = 140$, $t = 381$, $\Delta' = 22$.
The velocity parameters in Eq.~(\ref{hblg}) are related to the $\gamma$
parameters by $v_i=a\gamma_i\sqrt{3}/2\hbar$, where $a=0.246$ nm is the
lattice constant.

The potentials are given in the figure captions. The subscript $l$ denotes
potentials on the left side of a step and the left and right sides of a
barrier, $r$ denotes the right side of a step and $c$ denotes the center of
a barrier.

\subsection{Potential steps in unbiased BLG}
\label{UnbiasedSection}
\begin{figure}
  \begin{center}
    \hspace{-4mm}
  \includegraphics[width=4.1cm, angle=-90]{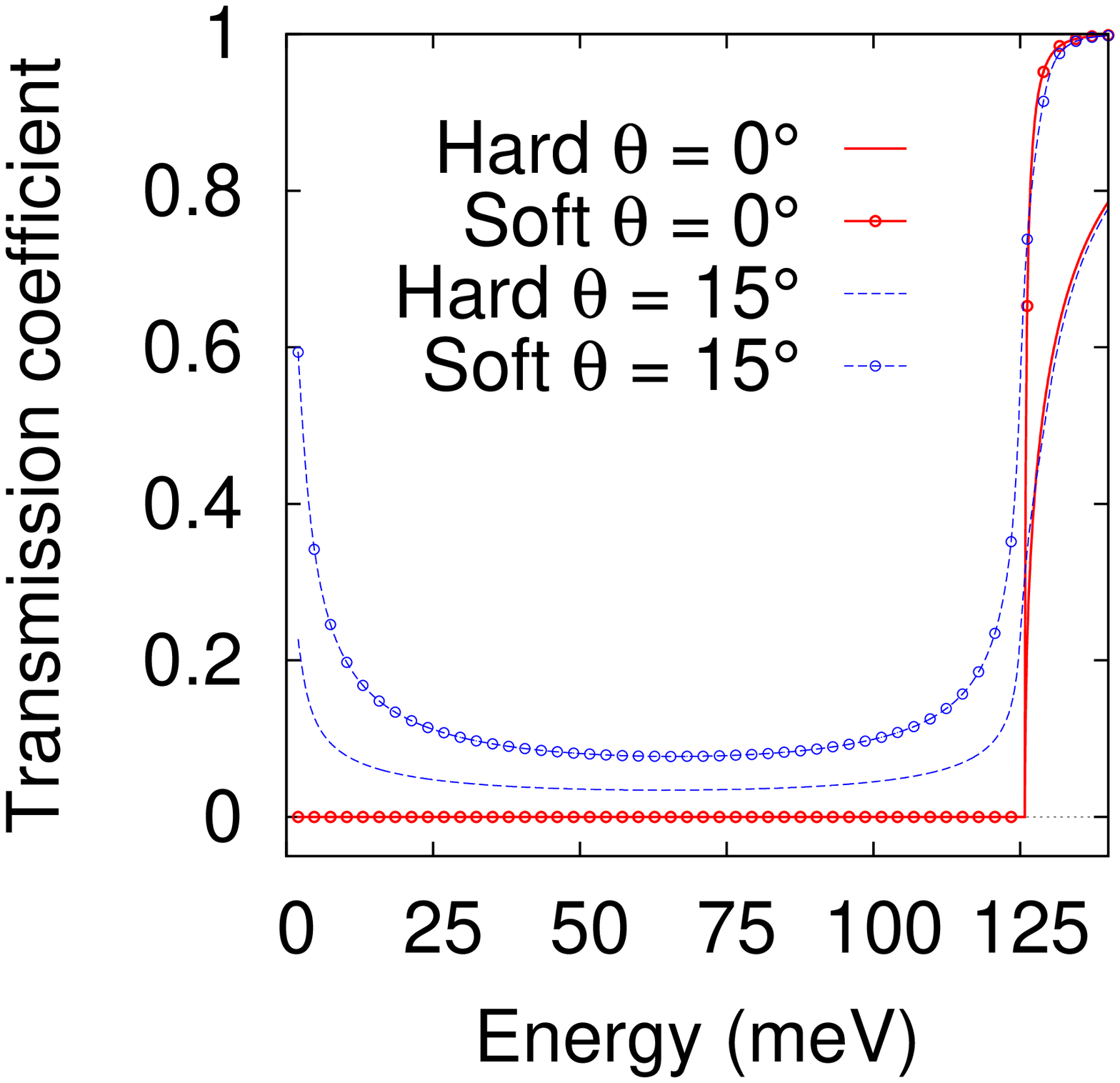}
    \hspace{3mm}
  \includegraphics[width=4.1cm, angle=-90]{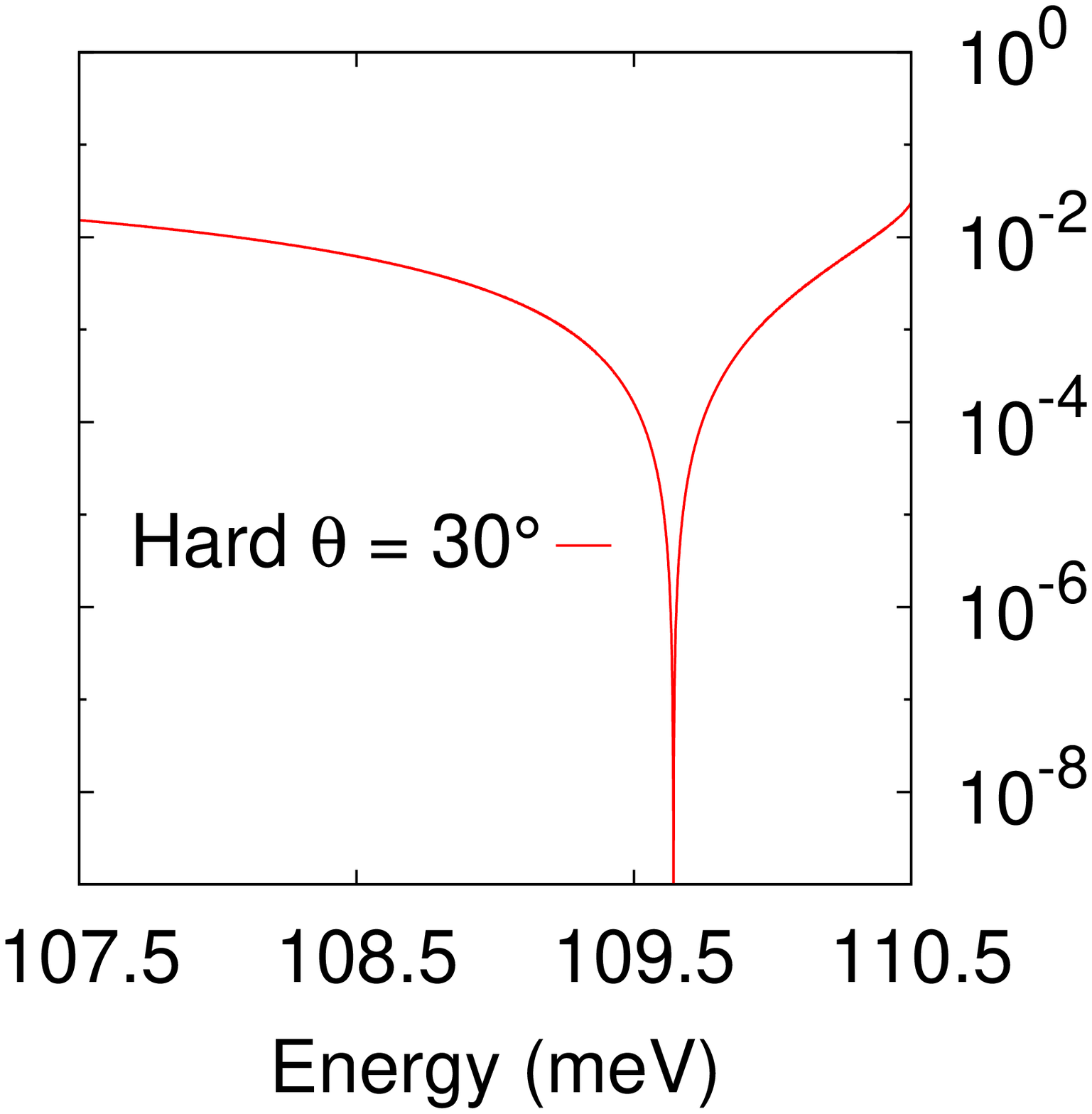}
  \caption{(Color online). Transmission coefficients for normal current
    incidence on potential steps of height 127 meV in unbiased BLG. Left:
    $\theta = 0^\circ$ armchair edge and $15^\circ$ midway edge,
    soft step width = 10 nm.
    Right: $\theta = 30^\circ$ zigzag edge.}
\label{normincfig}
\end{center}
\end{figure}

In unbiased BLG we have found AK tunneling only when the step edge is
parallel to an armchair direction or a zigzag direction.  In the armchair
case, AK tunneling occurs only when the incident current is normal to the
step edge (Section \ref{ArmchairSection}) and results from the swap
invariance of the 4-component Hamiltonian (Section \ref{SymmetrySection}).
This is the only case where AK tunneling occurs in an extended energy
range.  In the zigzag case, it may occur at normal or oblique current
incidence but only at critical values of the energy or angle of incidence
(Section \ref{ZigzagSection}).

\subsubsection{Armchair edge}
\label{ArmchairSection}

Fig.~\ref{normincfig} (left) shows transmission coefficients for normal
incidence on a potential step in unbiased BLG. The armchair directions
correspond to $\theta = n \pi / 3$, where $n$ is an integer; the $\theta =
0^\circ$ case is shown in the figure.  AK tunneling occurs in the energy
range where the incident state on the left side of the step is in the
conduction band and the transmitted state on the right side is in the
valence band. This range starts about 1-2 meV above the bottom of the
potential step and ends about 1-2 meV below the top. These energy offsets
occur because the conduction and valence bands overlap in a small energy
range when TW is present \cite{Partoens06,McCann07}. Except for the
offsets, AK tunneling at $\theta = 0^\circ$ is similar to that found
earlier for a hard step in the 2-component approximation without TW
\cite{Katsnelson06}. However in the presence of TW the occurrence of AK
tunneling depends strongly on the step orientation.

This is illustrated by the case of $\theta = 15^\circ$. This value of $\theta$
is midway between the $\theta = 0^\circ$ armchair direction and the $\theta
= 30^\circ$ zigzag direction. Fig.~\ref{normincfig} shows that in this case
zero transmission does not occur but it is still possible to observe a
large decrease in $T$ in the energy range between the conduction band edge
on the left and the valence band edge on the right.

Fig.~\ref{normincfig} (left) also shows that AK tunneling occurs
at both hard and soft steps. The soft step potential is $(V_0/2)(1 +
\tanh(x'/w))$, where $V_0$ is the step height and $w$ is the step width.
The conditions needed for AK tunneling at this soft step are
exactly the same as those needed for a hard step. When $\theta
= 15^\circ$, the large decrease in $T$ also occurs.

\subsubsection{Swap symmetry of Hamiltonian}
\label{SymmetrySection}

The AK tunneling at normal incidence occurs because when $k_y = 0$, the
4-component Hamiltonian for unbiased BLG is swap invariant and so is
the  coefficient matrix in Eq.~(\ref{eigkx}).
The swap operation is performed by the operator
\begin{equation}
  S = 
    \left( \begin{array}{cc}
    0 & \sigma_x \\
    \sigma_x & 0 \\
    \end{array} \right), \label{h2cblg}
\end{equation}
where $\sigma_x$ is a Pauli matrix and the zeros denote $2\times 2$
matrices whose elements are all zero. The eigenvalues of $S$ are $s = \pm
1$ and both are doubly degenerate. By expressing the Hamiltonian in the
basis formed by the eigenvectors of $S$ it can be shown that $\mathbf{e}_3$
and $\mathbf{e}_4$ are in the $s=-1$ subspace when $E$ is in the valence
band and in the $s= +1$ subspace when $E$ is in the conduction band.
Further, Eq.~(\ref{ferel}) shows that the same is true for
$\mathbf{f}_3^\dagger$.

The AK tunneling at a hard, armchair step in unbiased BLG is a consequence
of the fact that the swap eigenvalues, $s_{3l}$ of $\mathbf{f}_{3l}$ and
$s_{3r}$ of $\mathbf{e}_{3r}$ are of opposite sign.
Because $S^2 = I$, the $4\times 4$ unit matrix,
$\mathbf{f}^\dagger_{3l}\cdot\mathbf{e}_{3r} = \mathbf{f}^\dagger_{3l} S^2
\cdot \mathbf{e}_{3r} = s_{3l}s_{3r}\mathbf{f}^\dagger_{3l} \cdot
\mathbf{e}_{3r}= -\mathbf{f}^\dagger_{3l} \cdot\mathbf{e}_{3r}$.
Hence $\mathbf{f}^\dagger_{3l} \cdot\mathbf{e}_{3r}$
vanishes. Thus AK tunneling at a hard
step occurs throughout the energy range where the incident state is in the
conduction band and the transmitted state is in the valence band or vice
versa.

The AK tunneling at a soft, armchair step in unbiased BLG is also a
consequence of the swap symmetry. $M_{33}$ gives the amplitude of the
$\mathbf{e}_3$ contribution to the state, $\psi_l$, on the left of a step
when the state on the right is $\mathbf{e}_{3r} \exp(ik_{3r} x')$. That is
$M_{33} = \mathbf{f}^\dagger_{3l}\cdot \psi_l$. But the state on the right is
in the $s= -1$ subspace and remains in this subspace for all $x'$ as the
two subspaces are decoupled because of the swap symmetry. Thus $\psi_l$ is
in the $s= -1$ subspace and $M_{33}$ vanishes because
$\mathbf{f}^\dagger_{3l}$ is in the $s= +1$ subspace. Hence the occurrence
of AK tunneling  is independent of the shape of the step potential, as can
be seen in Fig.~\ref{normincfig} (left).

Another important consequence of swap symmetry is that complete evanescent
to propagating mode conversion occurs at armchair potential steps in
unbiased BLG. The propagating states in the conduction band have opposite
swap symmetry to those in the valence band and the same is true for the
evanescent states. The same analysis that led to $M_{33} = 0$ then shows
that the propagating-propagating and evanescent-evanescent elements of the
transfer matrix vanish, that is
$M_{11} = M_{12} = M_{21} = M_{22} = M_{33} = M_{34} = M_{43} = M_{44} = 0$.
Hence any propagating state on one side of a step must couple to an
evanescent state on the other side.

\subsubsection{Zigzag edge}
\label{ZigzagSection}
\begin{figure}
  \begin{center}
    \includegraphics[width=4.1cm, angle=-90]{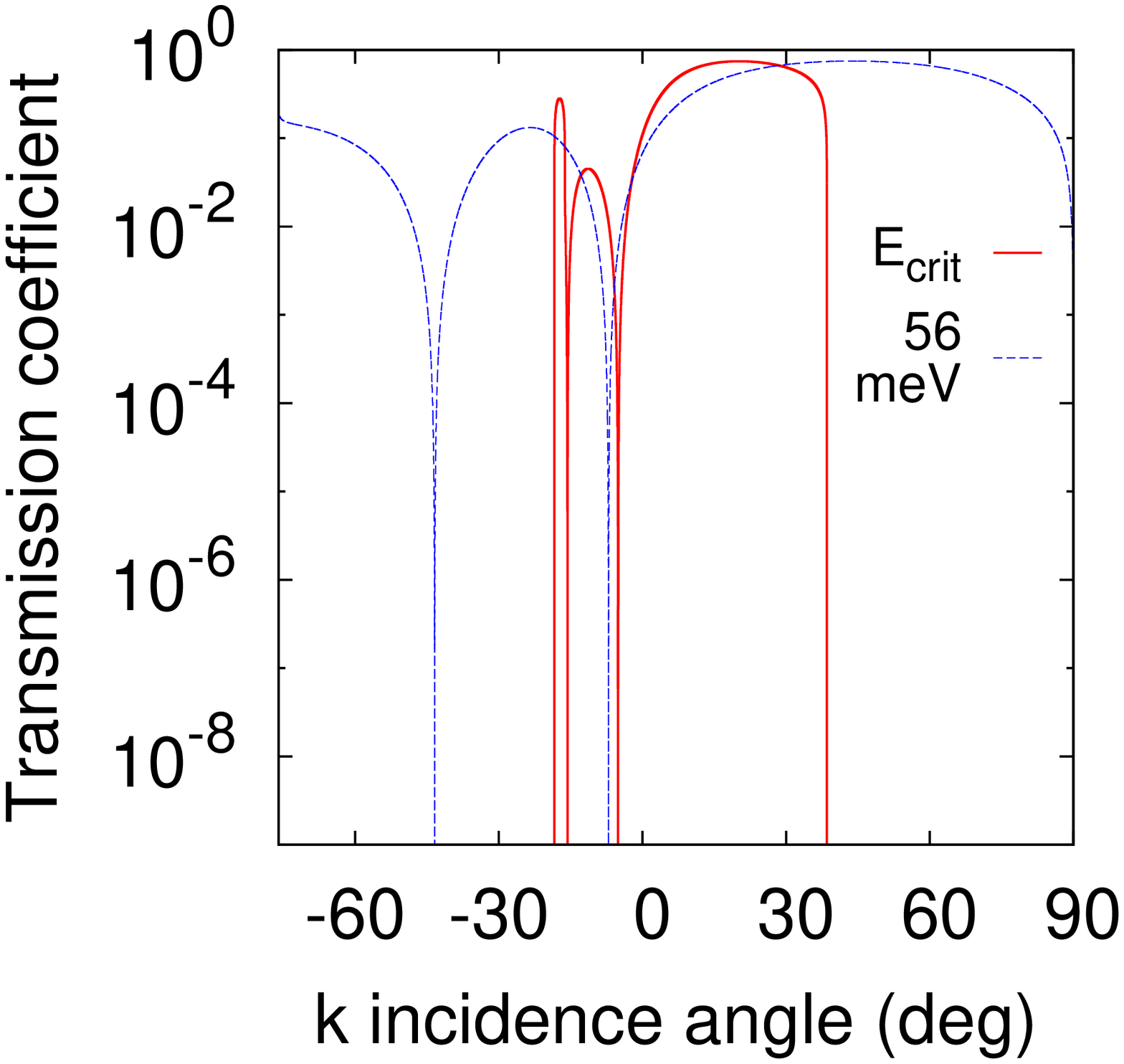}
    \includegraphics[width=4.1cm, angle=-90]{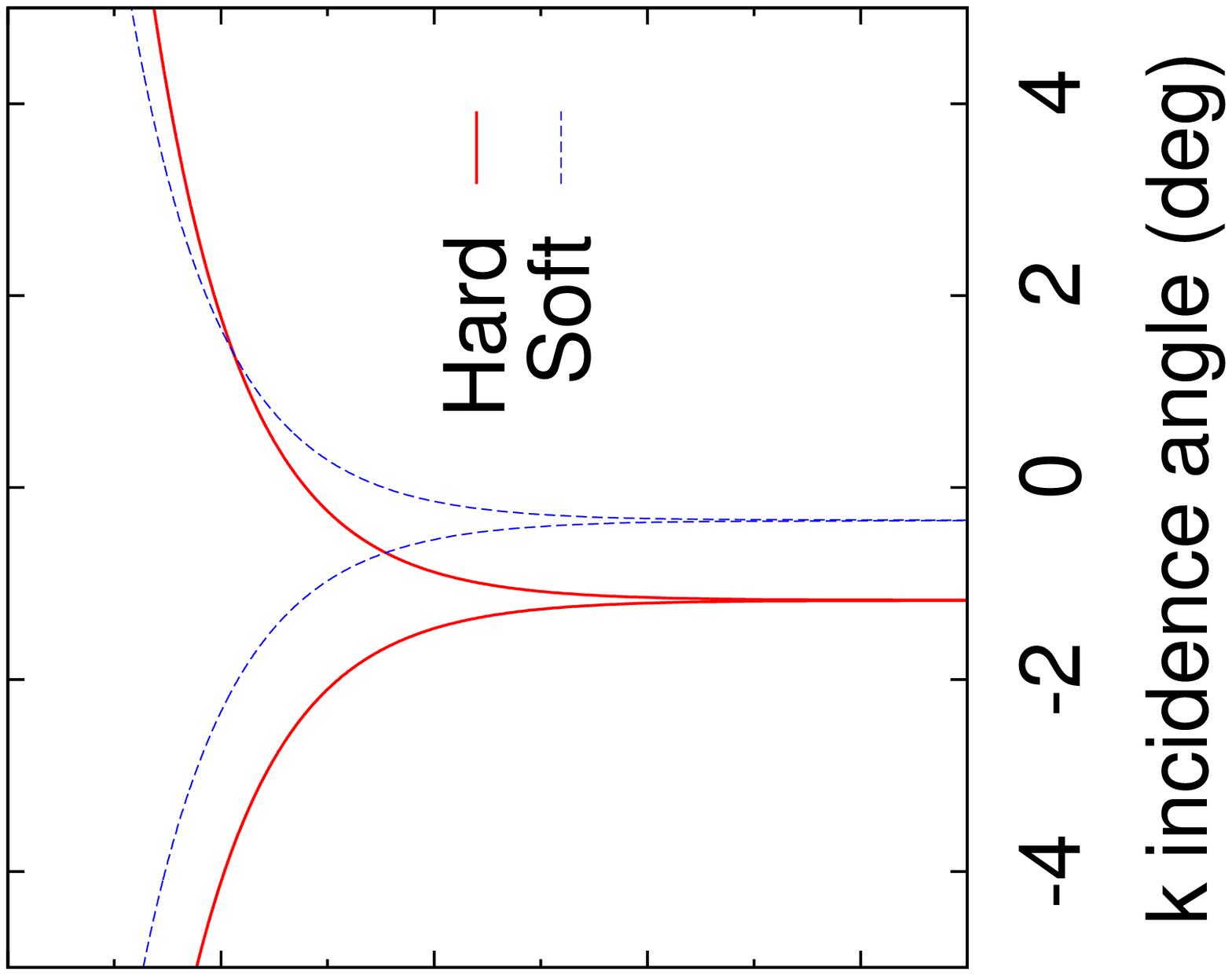}
  \caption{(Color online). Transmission coefficients for oblique
    $\mathbf{k}$ incidence
    on potential steps. Left: unbiased BLG for the same hard step potential
    and zigzag edge as in Fig.~\ref{normincfig} (right). AK zeros occur at
    $\phi_k \sim -43.41^\circ$, $\phi_c \sim -64.22^\circ$ and
    $\phi_k \sim -7.093^\circ$, $\phi_c \sim 18.57^\circ$ for $E = 56$ meV;
    $\phi_k \sim -15.66^\circ$, $\phi_c \sim 0^\circ$ and
    $\phi_k \sim -5.082^\circ$, $\phi_c \sim 14.74^\circ$
    for $E = E_{crit} \sim 109.6$ meV. At this energy, mode 3 is
    propagating when $-18.41 \alt \phi_k \alt -16.21^\circ$.
    Right: biased BLG near the $60^\circ$ armchair edge,
    $V_{1l} = -14$ meV, $V_{2l} = +14$ meV,
    $V_{1r} = 146$ meV, $V_{2r} = 108$ meV, $E = 56$ meV,
    $\theta \sim 56.54^\circ$ (hard step), $\theta \sim 60.09^\circ$
    (soft step, width = 10 nm). AK zeros occur at
    $\phi_k \sim -1.174^\circ$, $\phi_c \sim 3.612^\circ$ (hard step)
    and $\phi_k \sim -0.3420^\circ$, $\phi_c \sim -0.08361^\circ$ 
    (soft step).}
\label{oblincfig}
\end{center}
\end{figure}

The zigzag edges correspond to $\theta = n\pi/6$ where $n$ is an odd
integer.  Fig.~\ref{normincfig} (right) shows that AK tunneling occurs at
normal current incidence on a $\theta = 30^\circ$ zigzag step at a critical
energy $E_{crit} \sim 109.6$ meV. And Fig.~\ref{oblincfig} (left) shows
that AK tunneling occurs at oblique incidence on the same step over a wide
range of energies. In both figures the AK transmission zeros are very sharp
but $T$ is $\alt 1$\% within a few meV or a few degrees of the
zeros. Thus each AK zero is surrounded by an observable transmission
minimum. The cut-offs in $T(\phi_k)$ near $|\phi_k| = 30^\circ$ at $E =
E_{crit}$ are caused by total external reflection \cite{Maksym21}.

The AK tunneling at oblique incidence results from TW.
Without TW, AK tunneling in unbiased BLG occurs only
at normal incidence because the unnormalized $\mathbf{e}_3$ vectors in this
case are
\begin{equation}
  \mathbf{e}_3 = (c(\lambda - k_y), ~1, ~b, ~b c(\lambda + k_y))^T,
  \label{evecform}
\end{equation}
where $k_3 = i \lambda$, $c = i\hbar (v_0 - b v_4)/(E-V)$ and
$b = +1$ in the conduction band and $-1$ in the valence band. By evaluating
$\mathbf{f}^\dagger_{3l}\cdot\mathbf{e}_{3r}$ with these vectors it can be shown
that AK tunneling only occurs at normal incidence as found in earlier work
\cite{Katsnelson06} in the 2-component approximation without the $\Delta$
and $\gamma_4$ terms. However in the presence of TW, the
$\mathbf{e}_3$ vectors no longer have the simple form given in
Eq.~(\ref{evecform}) and AK tunneling occurs at oblique incidence as shown
in Fig.~\ref{oblincfig} (left).

The AK tunneling at normal current incidence shown in Fig.~\ref{normincfig}
(right) occurs at a critical condition when one of the AK transmission
zeros occurs exactly at a $\phi_k$ value that makes $\phi_c$ zero. As shown
in Fig.~\ref{oblincfig} (left), the AK zeros move to smaller
$|\phi_k|$ when the energy increases. When $E=E_{crit}$, an AK zero occurs at
$\phi_k \sim -15.66^\circ$, the $\phi_k$ value that makes the incident
current normal to the step edge. This results in the AK zero shown in
Fig.~\ref{normincfig} (right) which also occurs at $E=E_{crit}$.

\subsection{Potential steps in biased BLG}
\label{BiasedSection}

In biased BLG, AK tunneling does not occur over an extended energy
range because the bias potential breaks the swap symmetry. Nevertheless
AK tunneling does occur at critical energies or angles of incidence where
$\mathbf{f}^\dagger_{3l}\cdot\mathbf{e}_{3r}$ vanishes. These energies and
angles depend on the step edge orientation and the bias field
configuration, that is whether the bias fields on opposite sides of
the step are parallel or anti-parallel.

For all step edge orientations other than zigzag, AK tunneling occurs at a
critical pair of $\theta$ and $\phi_k$ values or a critical pair of
$\theta$ and $E$ values. A pair of values is needed because
$\mathbf{f}^\dagger_{3l}\cdot\mathbf{e}_{3r}$ is complex unless the step edge
orientation is zigzag. This means two parameters must be varied to ensure
that the  real and imaginary parts of
$\mathbf{f}^\dagger_{3l}\cdot\mathbf{e}_{3r}$ are both zero. We find these
zeros by fixing $E$ and varying $\theta$ and $\phi_k$.

Fig.~\ref{oblincfig} (right) shows an example of an AK transmission zero
which occurs at a hard step close to the $60^\circ$ armchair direction. The
form of $T(\phi_k)$ is similar to the form found in unbiased BLG
(Fig.~\ref{oblincfig} (left)) and $T(\phi_k)$ is again small within a few
degrees of the exact zero. The figure also shows an AK transmission zero
at a soft step close to the $60^\circ$ armchair direction. The positions of
the zeros in biased BKG depend on the step width because the swap symmetry
is broken. However in the example shown in Fig.~\ref{oblincfig} (right),
$\theta$ and $\phi_k$ only change by a few degrees when
the step wall is changed from hard to soft.

The bias fields in the case of Fig.~\ref{oblincfig} (right) are in the
anti-parallel configuration. Similar AK transmission zeros occur in the
parallel field configuration. However their position is more sensitive to
the bias field magnitude: when the magnitude increases from zero they move
away from the armchair direction rapidly.

AK tunneling also occurs in biased BLG when the step edge is parallel to a
zigzag direction. These directions are special because
$\mathbf{f}^\dagger_{3l}\cdot\mathbf{e}_{3r}$ is real. Then zeros can be found
by varying one parameter; we vary either $E$ or $\phi_k$. The resulting
form of $T$ is very similar to that found in unbiased BLG: typically there
are two zeros in $T(\phi_k)$ and there is a critical energy where a
transmission zero occurs at normal current incidence.

The occurrence of these zeros depends on the bias field configuration.  In
the anti-parallel case they occur at normal and oblique incidence with and
without TW up to at least $\simeq \pm 21$ meV bias. In the case of parallel
fields and oblique incidence they also occur up to at least $\simeq \pm 21$
meV bias when there is no TW. But if TW is present the bias magnitude must be
$\alt 14$ meV. In the case of parallel fields and normal incidence,
we have not found any AK zeros without TW and when TW is present the bias
magnitude must be $\alt 7$ meV.

\subsection{Potential barriers}
\label{BarrierSection}
\begin{figure}
  \begin{center}
  \includegraphics[width=4.1cm, angle=-90]{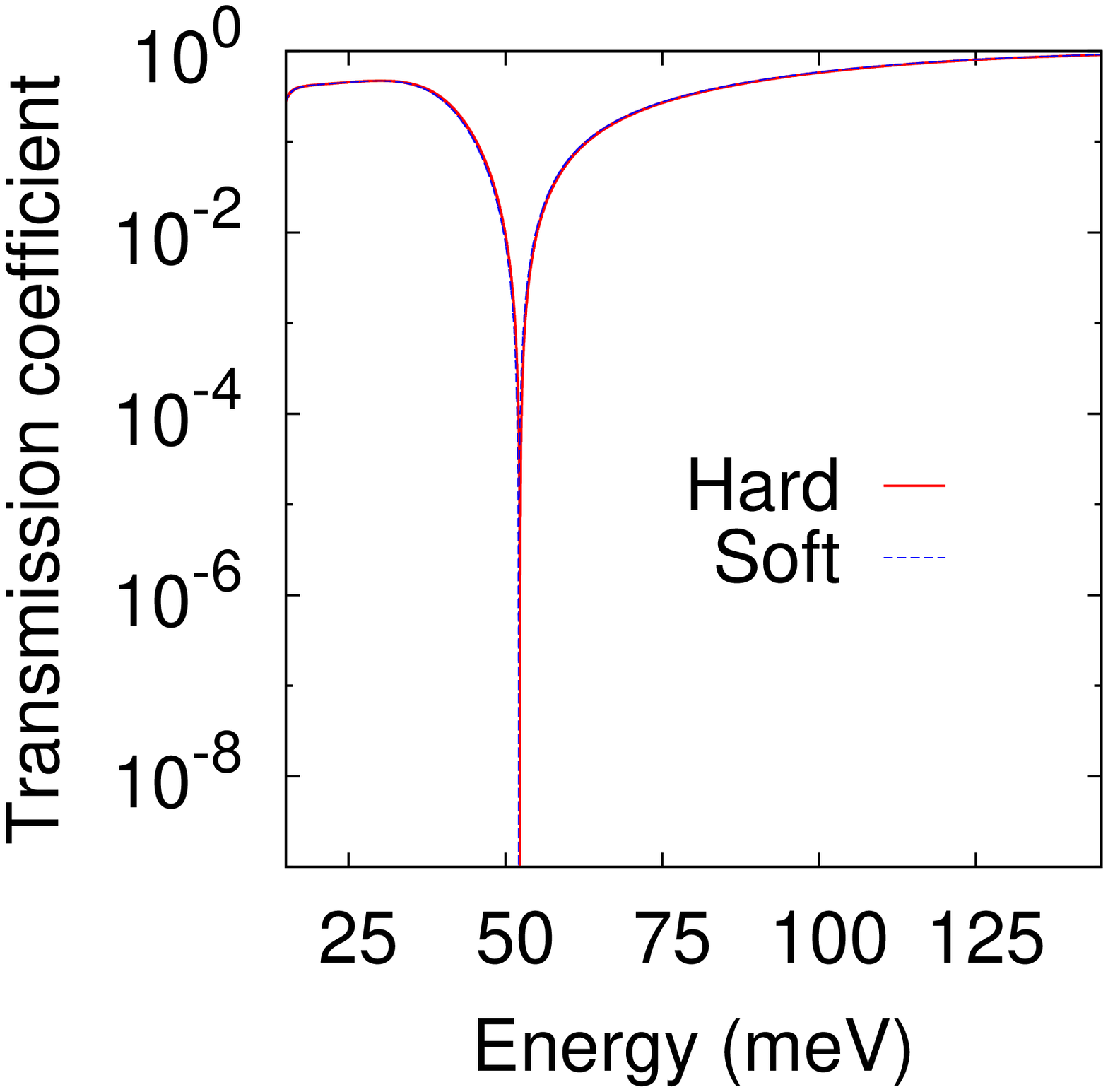}
  \hspace{2mm}\includegraphics[width=4.1cm, angle=-90]{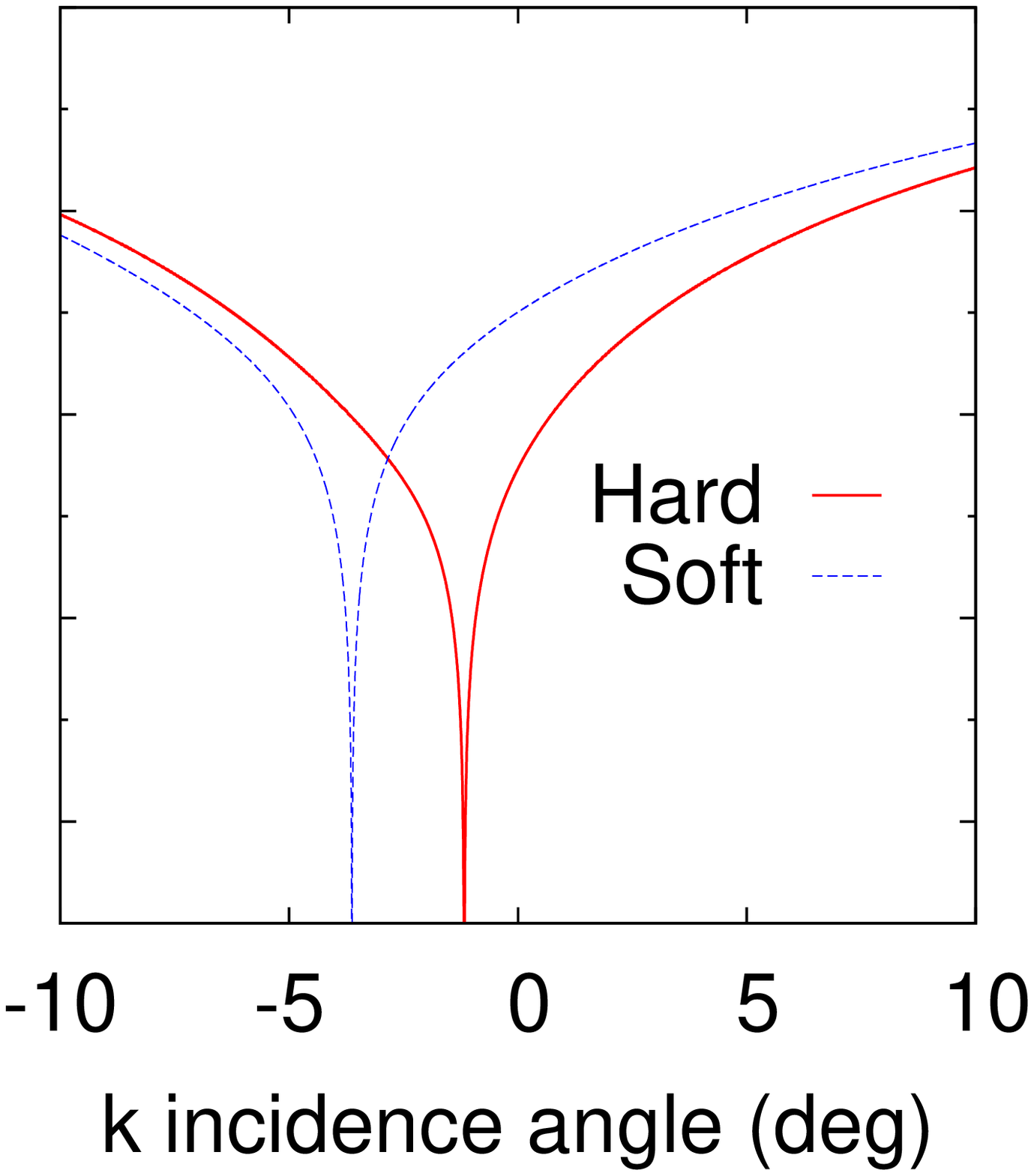}
  \caption{(Color online). Transmission coefficients for
    potential barriers in biased BLG, barrier width $\sim 8.913$ nm,
     soft wall width = 0.5 nm,
     $V_{1l} = -14$ meV, $V_{2l} = +14$ meV, $V_{1c} \sim 52.96$ meV,
     $V_{2c} \sim 103.3$ meV. Left: $k_{y'} = 0$ (normal $\mathbf{k}$ incidence),
     $\theta =30^\circ$ (zigzag edge). AK zeros occur at
     $E \sim 52.33$ meV, $\phi_c \sim 25.72^\circ$ (hard wall) and
     $E \sim 52.09$ meV, $\phi_c \sim 25.76^\circ$ (soft wall).
     Right: oblique $\mathbf{k}$ incidence, $E = 56$ meV,
     $\theta \sim 56.54^\circ$ (hard wall),
    $\theta \sim 52.48^\circ$ (soft wall). AK zeros occur at
    $\phi_k \sim -1.174^\circ$, $\phi_c \sim 3.612^\circ$ (hard wall)
    and $\phi_k \sim -3.628^\circ$, $\phi_c \sim 7.654^\circ$
    (soft wall).}
\label{barrierfig}
\end{center}
\end{figure}

AK tunneling occurs at potential barriers as well as steps. We show this
first for a barrier with hard walls. To find the necessary barrier width
and potential we set $E=56$ meV, $\theta \sim 56.54^\circ$ and
$\phi_k\sim-1.174^\circ$ as in Fig. \ref{oblincfig} and vary the barrier
width and $V_{1c}$ to find zeros of $M_{33}$. The barrier width that makes
$M_{33}$ zero also depends on $V_{2c}$; a width of $\simeq 9$ nm is obtained
with $V_{2c} \sim 103.3$ meV. The potential and barrier width found in this
way are used to compute the transmission coefficients in both parts of
Fig.~\ref{barrierfig}. AK tunneling occurs at normal $\mathbf{k}$
incidence when $\theta = 30^\circ$ and oblique $\mathbf{k}$ incidence
when $\theta \sim 56.54^\circ$.

Fig.~\ref{barrierfig} also shows that AK tunneling occurs at soft
potential barriers. The wall width is chosen to be slightly less than
an order of magnitude smaller than the barrier width. Nevertheless,
the position of the AK zero at oblique incidence is very
sensitive to the soft wall width.

The smallness of the barrier width is quite remarkable. The width is only
$\simeq 9$ nm yet tunneling through the barrier is blocked completely. AK
tunneling also occurs at wider barriers. When the edge is parallel to the
$30^\circ$ zigzag direction, we have found it at barriers up to about
$150$ nm wide in biased BLG and about $25$ nm wide in unbiased BLG,
see Fig. \ref{symfig} (d).

The possibility of AK tunneling at finite width barriers has been
mentioned in ref. \cite{Park12} on the basis of calculations in the
2-component approximation with TW for a barrier in unbiased BLG with the
edge parallel to an armchair direction. We have not found AK tunneling in
this case, both in the 4-component theory and the 2-component
approximation. The most likely cause of this discrepancy is that evanescent
waves are not taken into account in ref. \cite{Park12}.

\section{Valley dependence of AK tunneling}
\label{ValleySection}
\begin{figure}
  \begin{center}
  \includegraphics[width=3.cm, angle=-90]{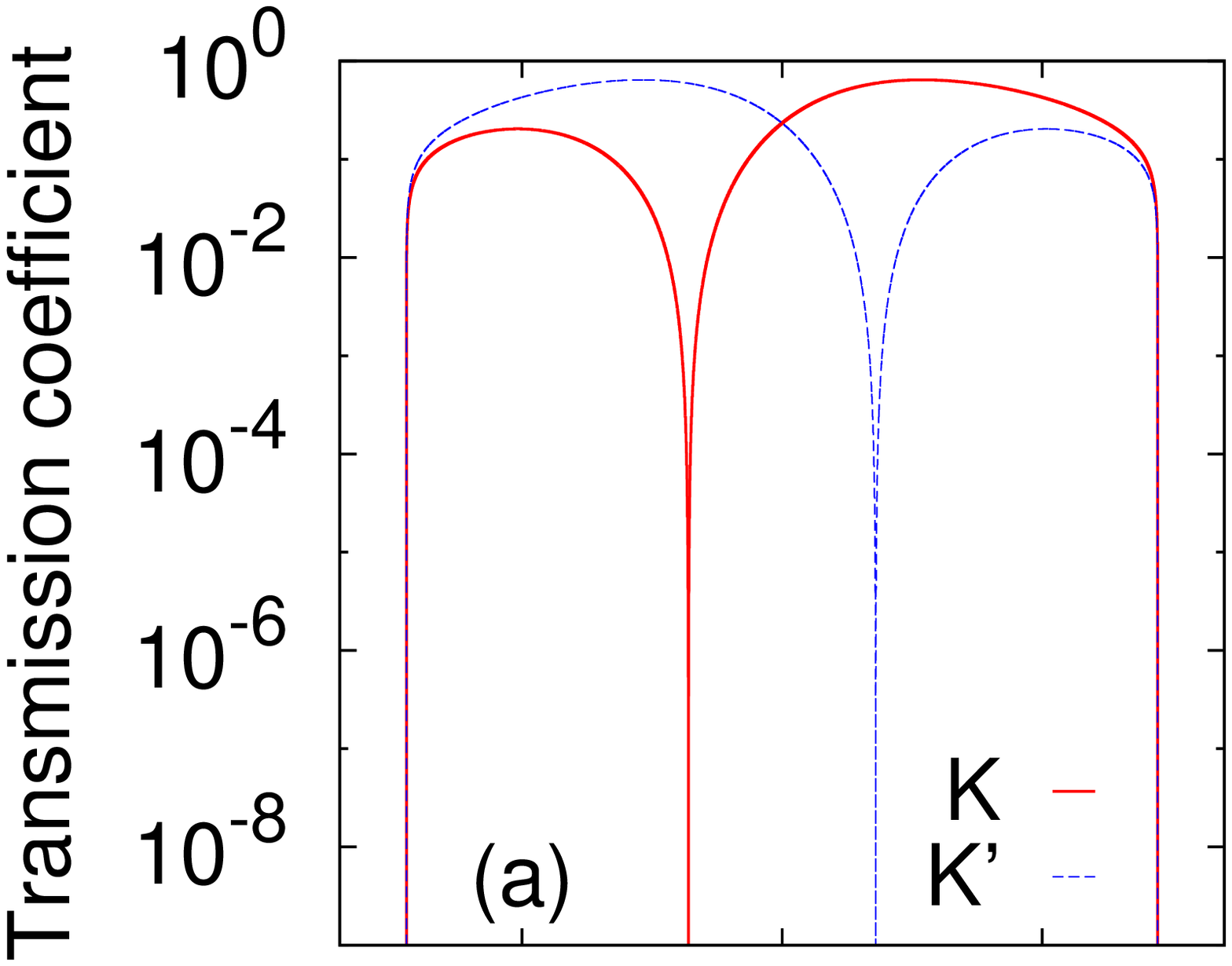}\hspace{1mm}
  \includegraphics[width=3.cm, angle=-90]{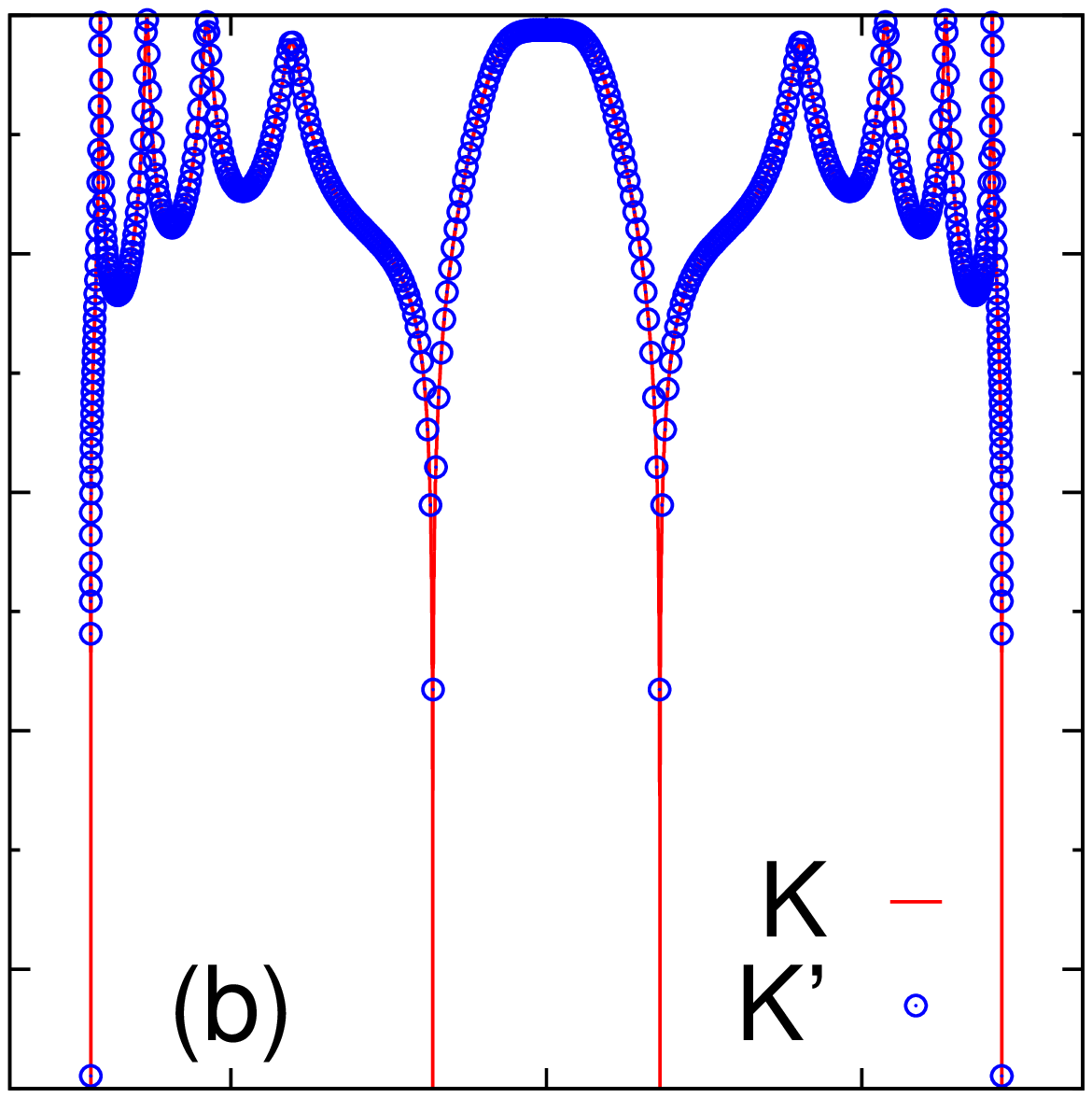}\\[1ex]

  % For some reason the widths for the bottom row figures have to be
  % increased to make all the printed figures the same size. bboxes are OK,
  % checked with gs.
  \includegraphics[width=3.815cm, angle=-90]{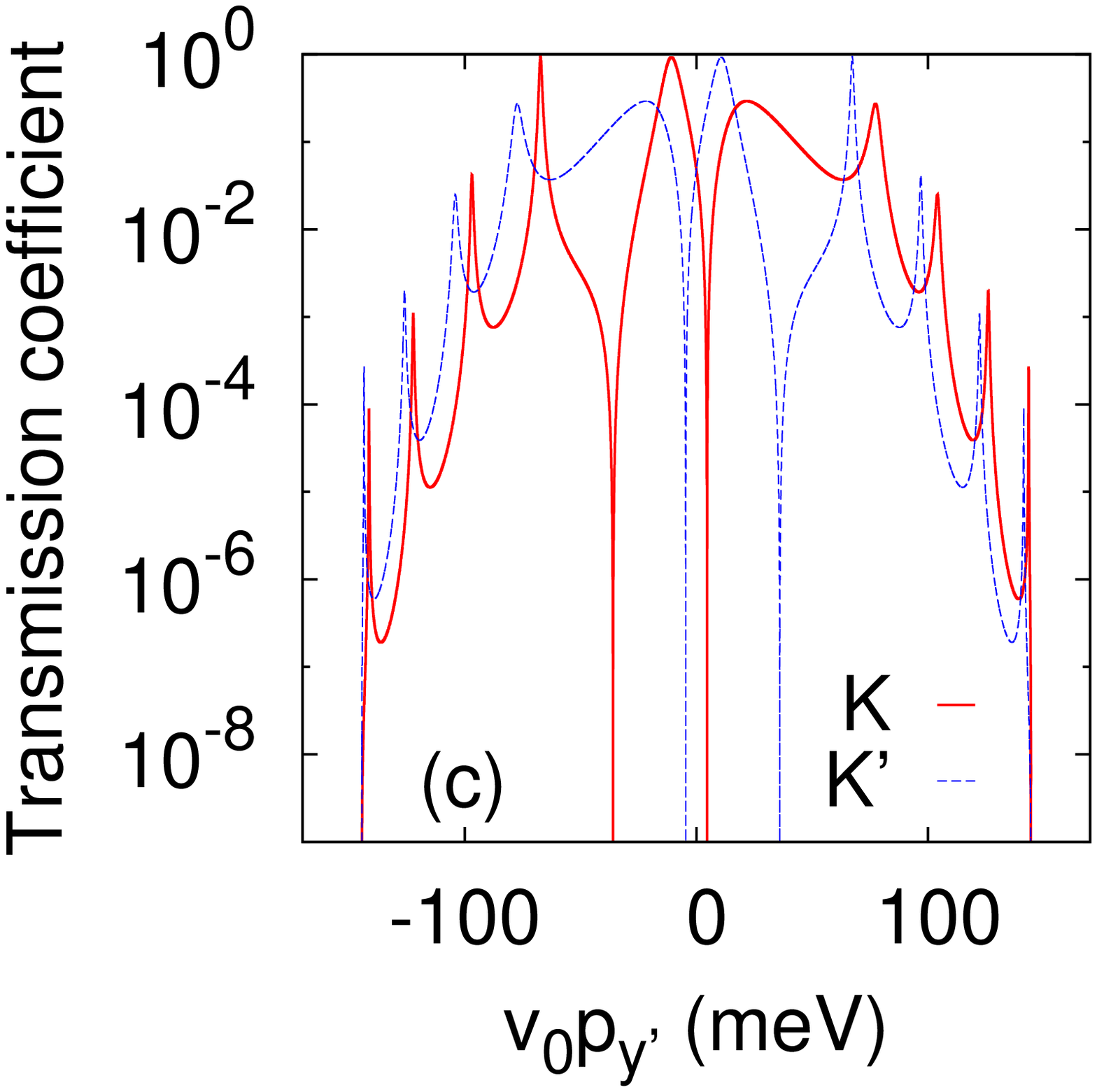}\hspace{1mm}
  \includegraphics[width=3.815cm, angle=-90]{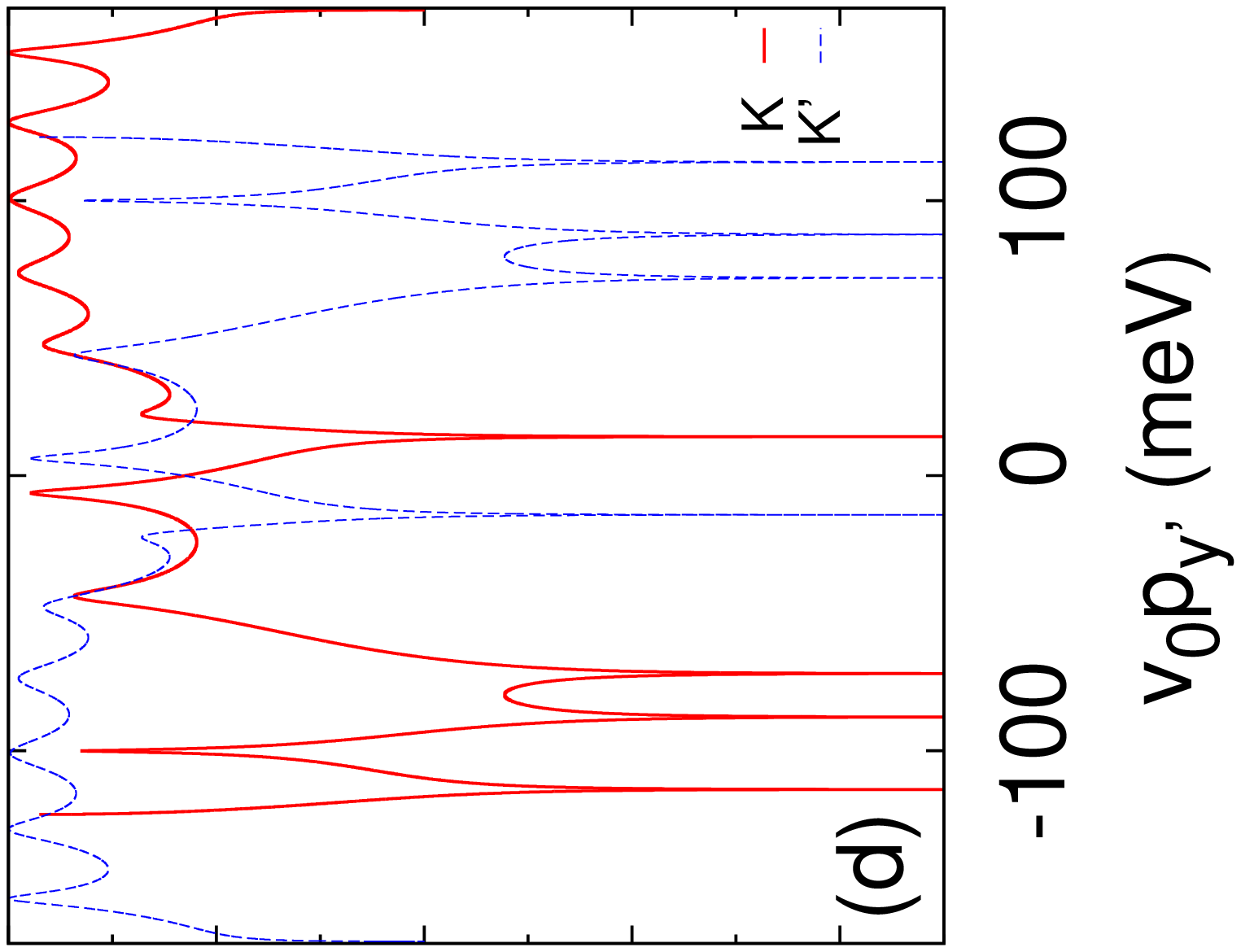}
  \caption{(Color online). Valley dependent AK tunneling at hard
    potential steps and barriers (width = 150 nm). $E = 56$ meV,
    $V_{1l} = 14$ meV, $V_{2l} = -14$ meV, $V_{1r} = V_{1c} = 146$ meV,
    $V_{2r} = V_{2c} = 108$ meV (parallel fields).
    (a) Valley asymmetric transmission at a step, no TW.  AK
    zeros at $v_0p_{y'} \sim \pm 35.99$ meV, $\phi_k = \phi_c \sim \pm 14.43^\circ$.
    (b) Valley symmetric transmission at a $x'$-inversion symmetric
    barrier, no TW.  AK zeros at $v_0p_{y'} \sim \pm 35.99$ meV, $\phi_k = \phi_c
    \sim \pm 14.43^\circ$.
    (c) Valley asymmetric transmission at a
    barrier with broken $x'$-inversion symmetry and no TW. See main text
    for details of symmetry breaking.  AK zeros at $v_0p_{y'} \sim \pm 35.99$
    meV, $\phi_k = \phi_c \sim \pm 14.43^\circ$; $v_0p_{y'} \sim \pm 4.594$ meV,
    $\phi_k = \phi_c \sim \pm 1.824^\circ$.
    (d) Valley asymmetric
    transmission at a $x'$-inversion symmetric barrier with TW;
    edges in $30^\circ$ zigzag direction.  AK zeros at $v_0p_{y'} \sim \pm 114.0$
    meV, $\phi_k \sim \pm 46.22^\circ$, $\phi_c \sim \pm 69.82^\circ$; $v_0p_{y'} \sim
    \pm 87.71$ meV, $\phi_k \sim \pm 31.22^\circ$, $\phi_c \sim \pm 33.59^\circ$;
    $v_0p_{y'} \sim \pm 71.91$ meV, $\phi_k \sim \pm 25.31^\circ$, $\phi_c \sim \pm
    16.41^\circ$; $v_0p_{y'} \sim \pm 14.19$ meV, $\phi_k \sim \pm 6.061^\circ$,
    $\phi_c \sim \pm 27.97^\circ$.  }
\label{symfig}
\end{center}
\end{figure}

The condition for AK tunneling can be valley-dependent because in BLG the
transmission coefficient can be valley-dependent. Because of time reversal,
the transmission coefficients in the two valleys satisfy
\begin{equation}
  T_K(k_{y'}) = T_{K'}(-k_{y'}),
  \label{timereveq}
\end{equation}
see Appendix \ref{SymmetryAppendix}. In principle, Eq.~(\ref{timereveq}) allows valley-dependent
transmission to occur. However, if $T(k_{y'}) = T(-k_{y'})$ within each
valley, Eq.~(\ref{timereveq}) gives $T_K(k_{y'}) = T_{K'}(k_{y'})$. Hence
valley-dependent transmission can occur only when the symmetry of
$T(k_{y'})$ is broken within each valley.

In BLG there are two symmetry breaking mechanisms.
The first is TW. This breaks the symmetry because the
constant energy contours are not symmetric in $k_{y'}$ unless the step
edge is parallel to an armchair direction.

The second mechanism is asymmetry of the potential, that is $V_i(x') \ne
V_i(-x')$. This allows valley asymmetric transmission even in the
absence of TW.

The transmission coefficient $T_K(k_{y'})$ for the potentials
$V_i(x')$ is related by symmetry to the transmission coefficient
$\hat{T}_{K'}(k_{y'})$ for the spatially inverted potentials $V_i(-x')$,
see Appendix \ref{SymmetryAppendix}.
In the presence of TW,
$ T_K(k_{y'}, \theta) = \hat{T}_{K'}(k_{y'}, \theta\pm \pi/3)$
but without TW
\begin{equation}
  T_K(k_{y'}) = \hat{T}_{K'}(k_{y'}).
  \label{xreveq}
\end{equation}
If the potentials are symmetric, $T = \hat{T}$ in each valley hence
$T(k_{y'})$ is valley symmetric. Otherwise $T(k_{y'})$ is in general valley
asymmetric. This counter-intuitive relation between the symmetry of $T$ in
the transverse direction and the symmetry of $V$ in the longitudinal
direction results from the fact that $\pi_{x'} = p_{x'}$ in the $K$-valley
and $-p_{x'}$ in the $K'$ valley.

Fig.~\ref{symfig} illustrates valley-dependent transmission in BLG. We plot
$T$ as a function of $v_0 p_{y'} = v_0 \hbar k_{y'}$ to show the valley
symmetry or asymmetry explicitly. The transmission coefficients without TW
are computed by setting $v_3 = 0$ and retaining all the other terms in the
Hamiltonian. Part (a) shows $T(k_{y'})$ for a potential step without
TW. The transmission is valley-dependent in accordance with
Eq.~(\ref{xreveq}) and $T_K(k_{y'}) = T_{K'}(-k_{y'})$ in accordance with
Eq.~(\ref{timereveq}).  Part (b) shows $T(k_{y'})$ for a potential barrier
without TW. The barrier potential is symmetric in $x'$ so the transmission
is symmetric in $k_{y'}$. Part (c) shows $T(k_{y'})$ for no TW and the same
potential barrier as for part (b) plus an additional potential that makes
the barrier asymmetric. The
transmission is valley-dependent in accordance with Eq.~(\ref{xreveq}).
(In each layer the symmetry breaking potential consists of
a constant shift applied in the $x'$ range $110 \le x' \le 150$ nm, where
the origin is at the entrance edge of the barrier and the barrier width is
150 nm. The shifts are -80 meV in layer 1 and -40 meV in layer 2.)
Part (d) shows $T(k_{y'})$ for the same symmetric potential barrier as for
part (b) but with TW. The transmission is valley-dependent and the
transmission coefficients satisfy Eq.~(\ref{timereveq}).

An important consequence of Eqs.~(\ref{timereveq}) and (\ref{xreveq}) is
that AK tunneling is valley-dependent and this can be seen in
Fig.~\ref{symfig}. If there is an AK zero at position $k_{y'}$ in a
particular valley, one also occurs at $-k_{y'}$ in the other valley. This
can result in a very large difference in the transmission coefficients in
the two valleys. For example, in part (d) near $v_0p_{y'} = \pm 80$ meV,
the transmission coefficients in the two valleys differ by over 4 orders of
magnitude. It should be possible to use this effect to realize a valley
polarizer, see Section \ref{ExperimentalSection}

The large valley dependence of the transmission does not occur in monolayer
graphene (MLG) at typical carrier energies. First, because TW is weak in
MLG unless the carrier energy is high \cite{Pereira09}. Secondly, because the
equivalent of the swap symmetry in MLG is site interchange, an operation
performed by $\sigma_x$. In each valley the MLG Hamiltonian satisfies
$\sigma_x H(k_{y'}) \sigma_x = H(-k_{y'})$. This has the consequence that
$T(k_{y'}) = T(-k_{y'})$ in each valley. Hence the potential asymmetry
mechanism is not available in MLG.

\section{Experimental consequences}
\label{ExperimentalSection}

The ideal arrangement for experimental investigation of the effects we
have reported is a potential barrier in the ballistic transport regime
\cite{Maksym21,Varlet14,Cobaleda14,Nam17,Oka19}. The barrier geometry has
the advantage that electrodes can be placed on the exit side to collect the
outgoing current while operation in the ballistic transport regime allows
the incidence conditions to be controlled. We envisage an arrangement
similar to the one suggested in our earlier work \cite{Maksym21} where a
collimated beam of electrons \cite{Barnard17} is incident on a potential
barrier formed by a top gate and a bottom gate is used to set the
Fermi level.

To obtain a clear signal, the incidence conditions should be set so that AK
tunneling occurs in both valleys. Eq.~(\ref{timereveq}) shows that this
requires $k_{y'} = 0$ as in Fig.~\ref{barrierfig}. It should be possible to
satisfy this condition experimentally by fixing the collimator position and
varying the gate voltages. Although the AK zeros are very sharp, we have
found that $T$ remains small, $\alt 1$ to 0.01\%, over a
measurable range of incidence parameters centered on the exact zero.  This
drop in $T$ is the experimental signature of AK tunneling. However when TW
is strong, several incident $\mathbf{k}$ states may carry current at the
same $\phi_c$ \cite{Maksym21a}. The $\phi_c$ ranges where this happens are
of small width, only $\simeq 0.4^\circ$, and should be avoided to obtain a
clear signal of AK tunneling.

The experimental arrangement we have suggested becomes a valley polarizer
when $k_{y'} \ne 0$. Then
if the collimator is aligned so that carriers are incident at the critical
angle for AK tunneling, transmission takes place only in one valley, while
carriers in the other valley are reflected away from the barrier. This
mechanism is similar to valley polarization by total external reflection
\cite{Maksym21} but can generate valley polarization even without
TW.

\section{Relation between 4-component and 2-component theories}
\label{2cSection}

In this section we show that the exact condition for AK tunneling in the
2-component approximation is simply the orthogonality condition,
Eq.~(\ref{antikcon}), with the exact 4-component polarization vectors
replaced with approximate ones (Section \ref{2cConditionSection}). We then
show that in the case of normal incidence this condition is equivalent to
the pseudospin conditions given by earlier authors \cite{Katsnelson06,
  Gu11, Park11, Park12}  (Section \ref{2cNormincSection}). Finally, we
compare transmission coefficients computed numerically with the 4-component
theory and the 2-component approximation (Section
\ref{2cNumericalSection}).

TW and other corrections were not taken into account in the first work on
AK tunneling in the 2-component approximation \cite{Katsnelson06,
Park11}. In this section we set $v_3$, $v_4$ and $\Delta'$ in
Eq.~(\ref{hblg}) to zero so that our 2-component Hamiltonian is the same as
in refs. \cite{Katsnelson06} and \cite{Park11}.

\subsubsection{Condition for AK tunneling in the 2-component approximation}
\label{2cConditionSection}

The 2-component approximation to the 4-component theory is obtained by
eliminating the dimer components, $\phi_{B1}$ and $\phi_{A2}$,
approximately \cite{McCann13}. To first order in $1/t$, the 2-component
state formed from the non-dimer components,
$(\tilde{\phi}_{A1}, \tilde{\phi}_{B2})^T$, is found from the
effective Hamiltonian
\begin{equation}
  \tilde{H}_K = -\frac{v_0^2}{t}
    \left( \begin{array}{cc}
    0 & (\pi_K^\dagger)^2 \\
    (\pi_K)^2 & 0 \\
    \end{array} \right) +
    \left( \begin{array}{cc}
    V_1 & 0 \\
    0 & V_2 \\
    \end{array} \right), \label{swapop4c}
\end{equation}
where tilde denotes the 2-component approximation. To the same order of
approximation, the dimer components satisfy
\begin{eqnarray}
  \tilde{\phi}_{B1} &=& -\frac{v_0}{t} \pi_K^\dagger \tilde{\phi}_{B2}
  \label{phib1eq}\\
  \tilde{\phi}_{A2} &=& -\frac{v_0}{t} \pi_K \tilde{\phi}_{A1}
  \label{phia2eq}.
\end{eqnarray}

The transmission and reflection coefficients may be found by imposing
appropriate boundary conditions at the step edge. As $\tilde{H}_K$ contains
second order derivatives, these conditions are continuity of each component
and its derivative \cite{Katsnelson06}. However this method of finding the
transmission and reflection coefficients obscures the relation between the
4-component theory and the 2-component approximation. We therefore
reformulate the 2-component approach so the boundary conditions become the
continuity of each component of an approximate 4-component state.

To do this we use the approximate dimer components given by
Eqs.~(\ref{phib1eq}) and (\ref{phia2eq}).
As the only $y'$-dependence is a factor of $\exp(ik_{y'}y')$,
Eqs.~(\ref{phib1eq}) and (\ref{phia2eq}) imply that the $x'$ derivatives of
$\tilde{\phi}_{B2}$ and $\tilde{\phi}_{A1}$ are continuous provided that
$\tilde{\phi}_{B1}$ and $\tilde{\phi}_{A2}$ are continuous. This allows the
derivative boundary condition to be replaced by a continuity condition on
the approximate 4-component state
$(\tilde{\phi}_{A1}, \tilde{\phi}_{B1},
\tilde{\phi}_{A2}, \tilde{\phi}_{B2})^T$. Next we show that the
corresponding approximate polarization vectors satisfy a biorthogonality
relation similar to Eq.~(\ref{biorthodef}).

Eqs.~(\ref{h2cblg}), (\ref{phib1eq}) and (\ref{phia2eq}) lead to an
eigenvalue equation for the approximate polarization vectors,
$\tilde{\mathbf{e}}_\alpha$,
\begin{equation}
  \tilde{v}_{x'K}^{-1}(\tilde{W} + p_{y'} \tilde{v}_{y'K})\tilde{\mathbf{e}}_\alpha
  = -\tilde{p}_\alpha \tilde{\mathbf{e}}_\alpha, \label{approxeigkx}
\end{equation}
where
\begin{equation}
  \tilde{W} = 
    \left( \begin{array}{cccc}
    V_1 - E& 0& 0 & 0\\
    0 & 0 & \hspace{5mm}t \hspace{5mm} & 0\\
    0 & \hspace{5mm} t \hspace{5mm} & 0 & 0\\
    0 & 0 & 0& V_2 - E \\
    \end{array} \right) \label{Vtildedef}
\end{equation}
and $\tilde{v}_{x'K}$ and $\tilde{v}_{y'K}$ respectively are
$v_{x'K}$ and $v_{y'K}$ with $v_3$ and $v_4$ set to zero. The matrix on the
left hand side of Eq.~(\ref{approxeigkx}) is again a general complex matrix
hence the approximate polarization vectors form a biorthogonal set. This
means biorthogonality can be used as described in Section \ref{4cTRSection}
to find the transmission coefficients in the 2-component approximation.
Thus the exact condition for AK tunneling in the 2-component approximation is
\begin{equation}
  \tilde{\mathbf{f}}^\dagger_{3l}\cdot\tilde{\mathbf{e}}_{3r} = 0,
  \label{approxantikcon}
\end{equation}
where $\tilde{\mathbf{f}}^\dagger_{3l}$ is an approximate left polarization
vector.

\subsubsection{Pseudospin conditions for AK tunneling at normal
incidence in the 2-component approximation}
\label{2cNormincSection}

In the case of normal incidence, Eq.~(\ref{approxantikcon}) leads to the
pseudospin conditions found by earlier authors \cite{Katsnelson06, Gu11,
  Park11, Park12}. We outline the proof of this here and give mathematical
details in the appendices.

At normal incidence in unbiased BLG, the approximate polarization vectors
are eigenvectors of the swap operator because $\tilde{v}_{x'K}$ and
$\tilde{W}$ in Eq.~(\ref{approxeigkx}) are swap symmetric. This means that
the condition for AK tunneling in the 2-component approximation is the
same as shown in Section \ref{SymmetrySection} for the 4-component theory.
This condition is equivalent to the pseudospin conservation condition
because the pseudospin eigenvalue of a 2-component polarization vector is
identical to the swap eigenvalue of the corresponding approximate
4-component vector (Appendix \ref{UnbiasedAppendix}).

In the case of biased BLG, the AK condition is that the expectation values
of the pseudospin on opposite sides of a step are the same. This condition
can be obtained by rotating the polarization vectors and using
Eq.~(\ref{approxantikcon}) to find the necessary rotation angle (Appendix
\ref{BiasedAppendix}).

\subsubsection{Numerical examples}
\label{2cNumericalSection}

In this section we present numerically computed transmission coefficients
for biased BLG and show that the critical energy and angle of incidence for
AK tunneling in the 2-component approximation may differ significantly
from those found in the 4-component theory.

\begin{figure}
  \begin{center}
  \includegraphics[width=4.3cm, angle=-90]{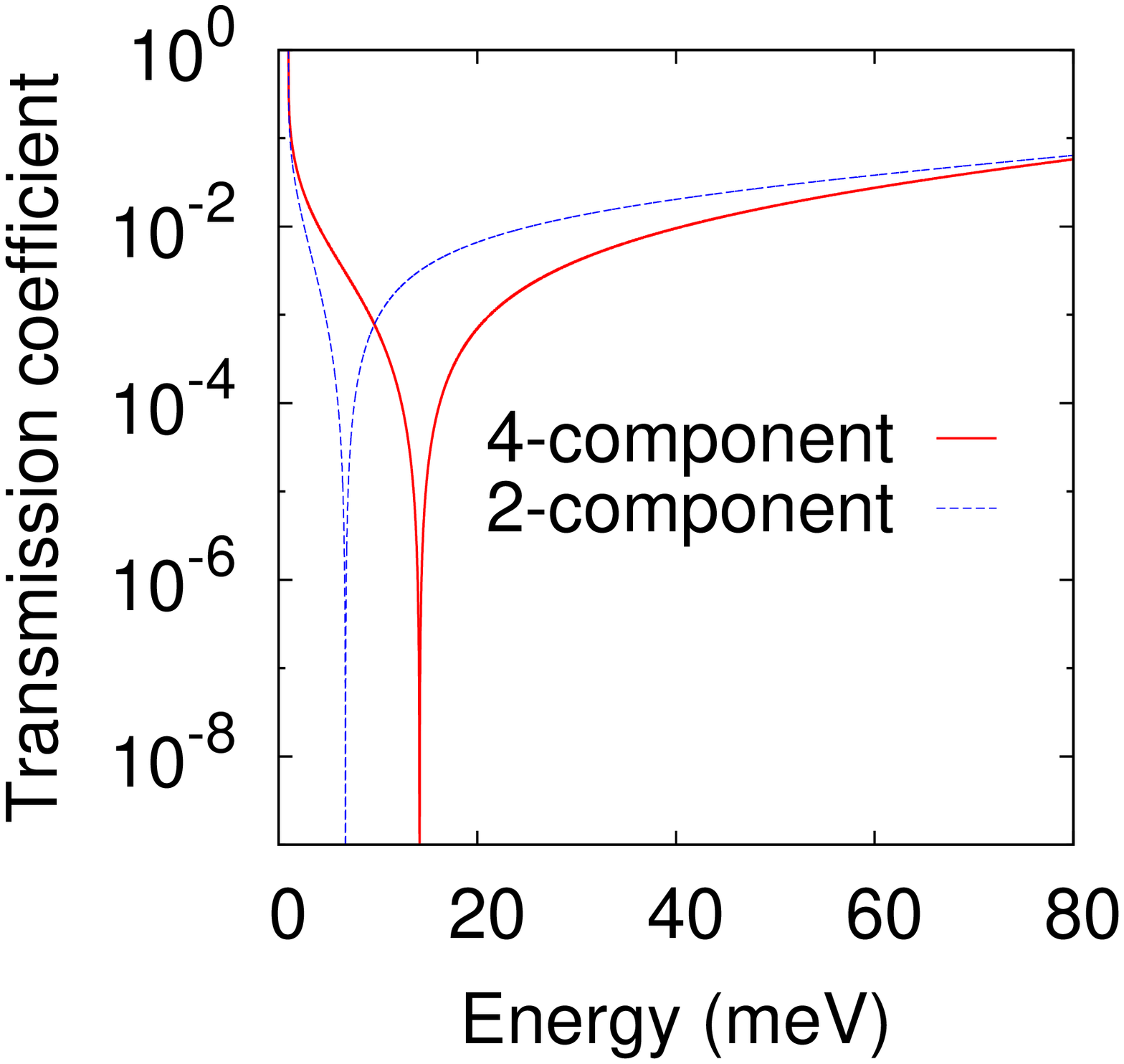}\hspace{1mm}
  \includegraphics[width=4.3cm, angle=-90]{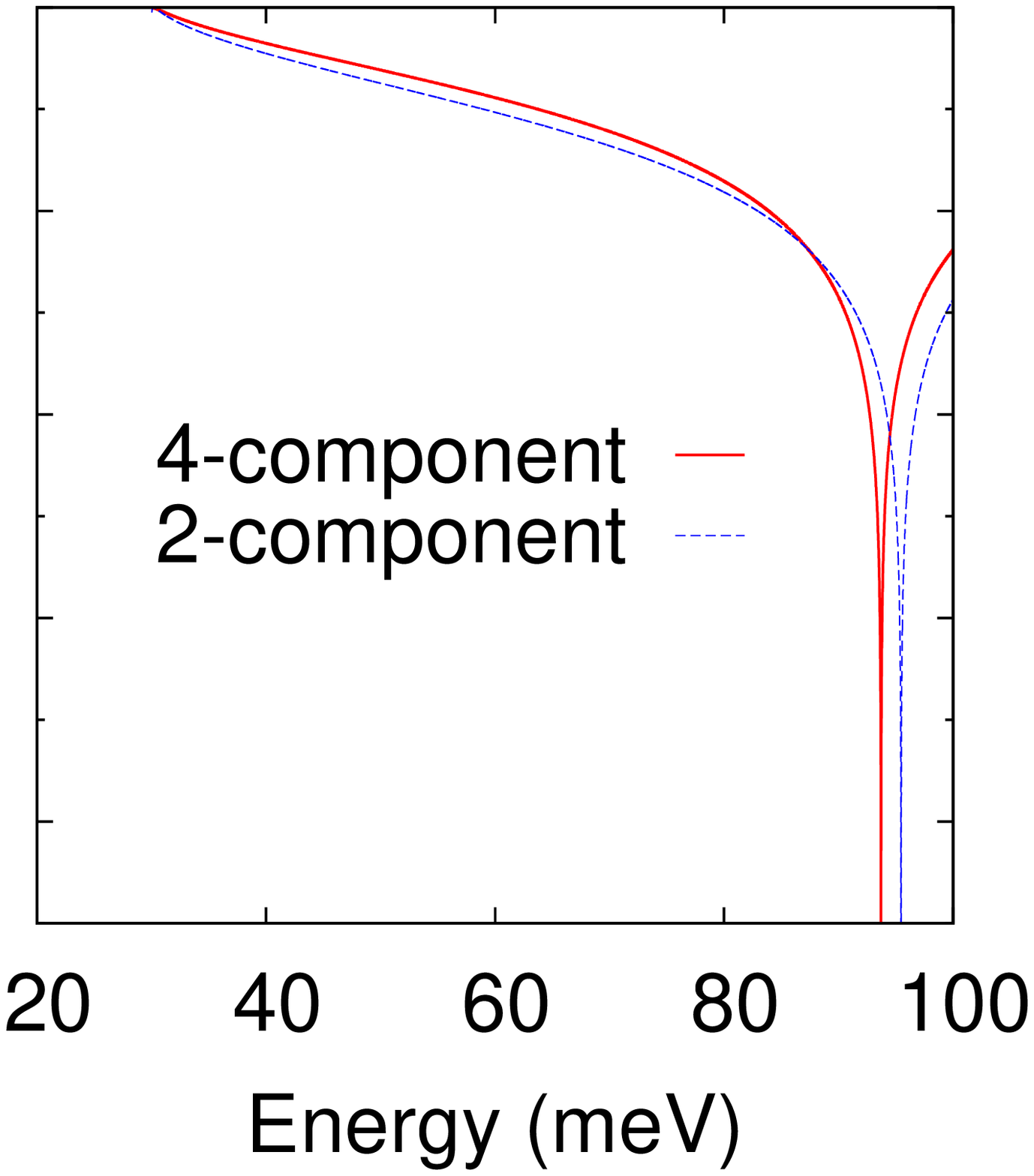}
  \caption{(Color online). Transmission coefficient for normal incidence on
    a hard step in BLG with $V_{1r} = 200$ meV,
    $V_{2r} = 150$ meV, no TW.
    Left: $V_{1l} = -1$ meV, $V_{2l} = +1$ meV. Right: $V_{1l} = -30$ meV,
    $V_{2l} = +30$ meV.}
\label{4c2cenfig}
\end{center}
\end{figure}

Fig.~\ref{4c2cenfig} shows transmission coefficients for electrons at
normal incidence. The critical energy for AK tunneling differs by about a
factor of 2 when there is a large bias mismatch. Then the 2- and 4-
component transmission coefficients near the critical energies differ by
one to two orders of magnitude (left side of figure).

\begin{figure}
  \begin{center}
  \includegraphics[width=4.3cm, angle=-90]{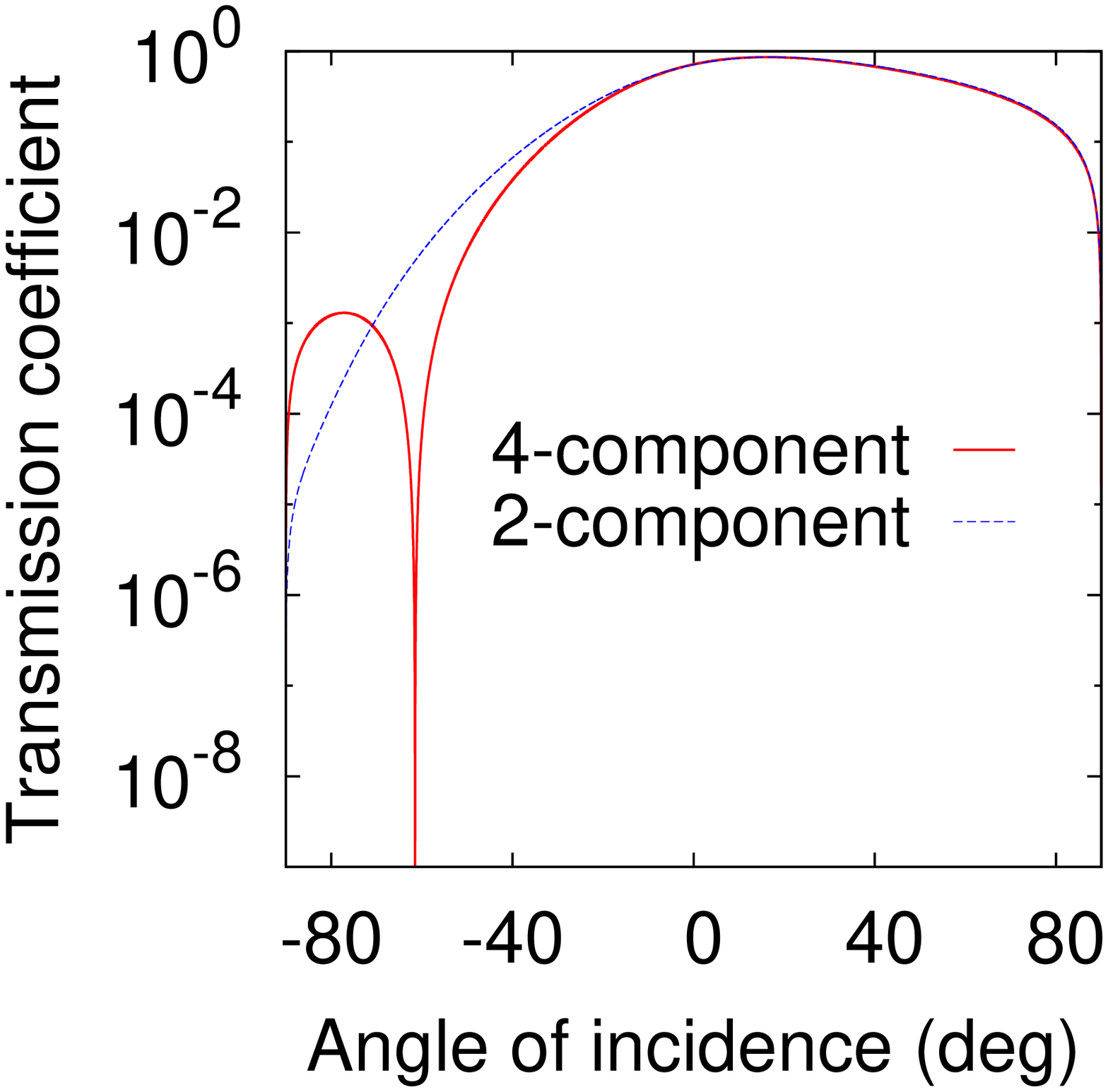}\hspace{1mm}
  \includegraphics[width=4.3cm, angle=-90]{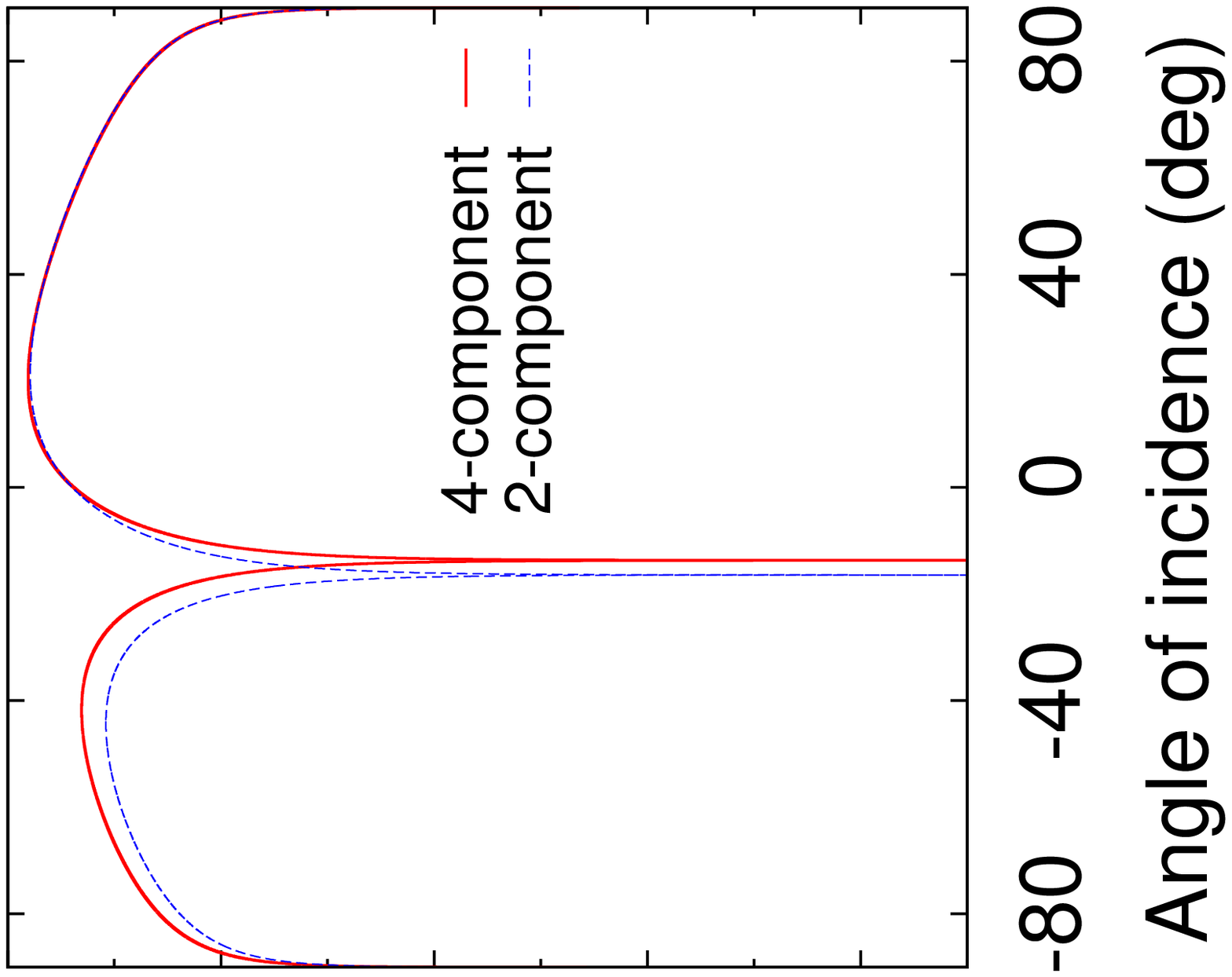}
  \caption{(Color online). Transmission coefficient for oblique incidence on
    a hard step in BLG with $V_{1l} = -14$ meV, $V_{2l} = +14$ meV,
    $V_{1r} = 146$ meV, $V_{2r} = 108$ meV, no TW.
   Left: $E = 16$ meV. Right: $E = 56$ meV.}
\label{4c2cphifig}
\end{center}
\end{figure}

Fig.~\ref{4c2cphifig} shows transmission coefficients for electrons at
oblique incidence. In this case AK tunneling in the 2-component
approximation has not been reported before but occurs in accordance with
Eq.~(\ref{approxantikcon}). But although AK zeros occur at both 16 meV
(Fig.~\ref{4c2cphifig}, left) and 56 meV (Fig.~\ref{4c2cphifig}, right)
in the the 4-component theory, there is no zero at 16 meV in the
2-component approximation. In general, the 2-component approximation
appears to be poor at large transverse momentum.

Figs.~\ref{4c2cenfig} and \ref{4c2cphifig} suggest that the reliability of
the 2-component approximation depends on the energy, angle of incidence and
interlayer bias. Because of this it is preferable to use the
4-component theory for numerical calculations. This requires no extra
computational cost or programming effort as the number of boundary
conditions is same in both cases.

\section{Summary and Conclusion}
\label{Discussion}
We have found exact conditions (Eqs.~(\ref{antikcon}) and
(\ref{mantikcon})) for AK tunneling in the 4-component continuum theory of
BLG. These conditions have 3 important consequences.

First, AK tunneling is ubiquitous but depends on the crystallographic
orientation of the step or barrier. In unbiased BLG at normal incidence on
a hard or soft armchair step it occurs because of the swap symmetry of the
4-component Hamiltonian. When swap symmetry is not present it occurs in
biased and unbiased BLG, not only at normal incidence but also at oblique
incidence on hard and soft steps and barriers with TW in all cases.

Secondly, AK tunneling at oblique incidence is
valley asymmetric provided that the transmission coefficient within each
valley is asymmetric in the transverse momentum. This asymmetry occurs
naturally because of TW but even without TW, asymmetry can be induced by
making the potential asymmetric in the longitudinal direction.

Thirdly, the exact condition for AK tunneling at normal and oblique
incidence in the 2-component approximation, Eq.~(\ref{approxantikcon}), is
just Eq.~(\ref{antikcon}) with the 4-component polarization vectors
replaced by approximate ones. At normal incidence
Eq.~(\ref{approxantikcon}) and swap symmetry lead to the pseudospin
conditions for AK tunneling in the 2-component approximation.  However,
there are cases where AK tunneling occurs in the 4-component theory but
not in the 2-component approximation.

The theoretical methods we have developed are applicable to analysis of
transmission and reflection in the tight binding approach, at least in the
case of normal incidence on a hard armchair step in unbiased BLG. We show
in Appendix \ref{TBAppendix} that in this case AK tunneling occurs as in
the continuum approach and that the transmission zero results from swap
symmetry and the orthogonality condition. Further investigation of
AK tunneling in the tight binding approach would require the
development of numerical methods to find all the $k_x$ values for a step of arbitrary
orientation and compute $T$ for a soft step.

Our findings are experimentally testable because we have shown that AK
tunneling occurs at experimentally realizable soft-walled potentials and
the transmission coefficient remains small over a measurable range centered
on the exact transmission zeros. It should be possible to observe AK
tunneling by using a graphene electron collimator \cite{Barnard17} coupled
to a potential barrier and working in the ballistic transport regime
\cite{Maksym21,Varlet14,Cobaleda14,Nam17,Oka19}. When this arrangement is
operated at zero transverse momentum it can detect AK tunneling and if it
is operated at non-zero transverse momentum, it functions as a valley
polarizer. The valley polarization is large and can be optimized by
adjusting the potential.

In summary, our work suggests that AK tunneling in BLG occurs under a wide
range of conditions, is experimentally detectable and can be used to make a
valley polarizer.

\begin{acknowledgments}
PAM thanks Prof. S. Tsuneyuki for hospitality at the Department of Physics,
University of Tokyo. The computations were done on the ALICE high performance
computing facility at the University of Leicester. HA is grateful for
support from the Core Research for Evolutional Science and Technology
``Topology" project from the Japan Science and Technology Agency (Grant No.
JPMJCR18T4) and JSPS KAKENHI Grant No. JP17H06138.
\end{acknowledgments}

\appendix
\section{Symmetry relations}
\label{SymmetryAppendix}
\subsection{General relations for steps and barriers}\
We have previously detailed some relations between transmission
coefficients for potential barriers \cite{Maksym21}. The only difference
between a barrier and a step is that the potential is the same on the
entrance and exit sides of the barrier, while for step it is different. All
of the relations we have already given can be generalized to the case of a
step. Here we state the relations that apply in the case when there is one
incident state and one propagating transmitted state.

All of the relations can be derived from the asymptotic $S$-matrix or the
Hamiltonian. The asymptotic $S$-matrix relates the amplitudes of the
incoming and outgoing waves in the asymptotic regime where the evanescent
wave amplitudes are negligible:
\begin{equation}
    \left( \begin{array}{c}
    r \\
    t \\
    \end{array} \right) =
    \left( \begin{array}{cc}
    S_{a} & S_{b}\\
    S_{c} & S_{d}\\
    \end{array} \right)
    \left( \begin{array}{c}
    i_0 \\
    x_0 \\
    \end{array} \right),  \label{sdef}
\end{equation}
where $i_0$ is the amplitude of the
incident wave, $r$ is the amplitude of the reflected wave, $t$ is the
amplitude of the transmitted wave and $x_0$ is the amplitude of a wave
incident from the right.

The relations \cite{Maksym21} between the $S$-matrix elements and
between the transmission coefficients are
\begin{eqnarray}
  |S_b| &=& |S_c|, \label{symrel1}\\
  T_K(k_{y'}, \theta) &=& T_{K'}(-k_{y'}, \theta), \label{symrel2}\\
  T_K(k_{y'}, \theta) &=& \hat{T}_{K'}(k_{y'}, \theta\pm \pi/3), \label{symrel3}\\
  T_K(k_{y'}, \theta) &=& T_{K'}(-k_{y'}, \pm\pi/3 - \theta), \label{symrel4}  
\end{eqnarray}
where $\hat{T}$ is the transmission coefficient for a barrier with the
spatially inverted potentials, $V_i(-x')$. Eq.~(\ref{symrel1}) is a
consequence of the unitarity of the $S$-matrix (or generalized unitarity
\cite{Maksym21} when the polarization vectors are not normalized to unit
flux). Eq.~(\ref{symrel2}) results from time reversal and
Eqs.~(\ref{symrel3}) and (\ref{symrel4}) occur because there are
transformations that relate the Hamiltonians at different values of $\theta$.

An additional relation occurs in the case of unbiased BLG because the swap
operator then transforms the Hamiltonian as
$SH(k_{y'}, \theta)S = H(-k_{y'}, -\theta)$. This leads to the relation
\begin{equation}
  T(k_{y'}, \theta) = T(-k_{y'}, -\theta), \label{symrel5}
\end{equation}
which holds in each valley.

\subsection{Relations between transmission coefficients for the 4 step
  configurations}
In Section \ref{ExamplesSection} we stated that the transmission
coefficients for the 4 step configurations in Fig.~\ref{stepconfigsfig} are
related. We detail these relations first for the case when there is one
incident state and one propagating transmitted state. This is the case for
all the transmission coefficients presented in the main text, except when
$-18.41 \alt \phi_k \alt -16.21^\circ$ in Fig.~\ref{oblincfig} (left). We
explain the changes that apply in this small range at the end of this
sub-section.

In the case of one incident state and one propagating transmitted state in
the presence of TW, all the transmission coefficients can be found from 2
independent functions of $k_{y'}$ and this reduces to 1 when the step edge
is parallel to an armchair direction or, when there is no bias,
a zigzag direction. Without TW only one function of $k_{y'}$ is needed.

Within each valley this is a consequence of Eq.~(\ref{symrel1}).
The physical meaning of $S_b$ and $S_c$ is that $S_c$
is the transmitted amplitude of a wave incident from the left and $S_d$ is
is the transmitted amplitude of a wave incident from the right. Then it
follows from Eq.~(\ref{symrel1}) that $T_{ru} = T_{lu}$ and $T_{rd} =
T_{ld}$, where the subscripts are defined in Fig.~\ref{stepconfigsfig}.
Once the transmission coefficients in one valley are known, those
in the other valley can be found from Eq.~(\ref{symrel2}). Thus only two
independent functions are needed to find all the transmission coefficients.
These functions can be taken to be $T_{lu}$ and $T_{ld}$. 

When the step edge is parallel to an armchair direction, only one
function is needed. In this case Eqs.~(\ref{symrel3}) and
(\ref{symrel2}) give $T_{luK}(k_{y'}, 0) = T_{ldK}(-k_{y'}, \pi / 3)$
while Eqs.~(\ref{symrel4}) and
(\ref{symrel2}) give $T_{ldK}(k_{y'}, 0) = T_{ldK}(k_{y'}, \pi / 3)$.
Hence $T_{ldK}(k_{y'}, 0) = T_{luK}(-k_{y'}, 0)$ and similarly
$T_{ldK}(k_{y'}, \pi/3) = T_{luK}(-k_{y'}, \pi/3)$. Thus only one function
is needed and can be taken to be $T_{lu}$.

When the step edge is parallel to a zigzag direction, similar reasoning
leads to $T_{ldK}(k_{y'}, \pi/6) = T_{luK}(-k_{y'}, \pi/2)$ and
$T_{ldK}(k_{y'}, \pi/2) = T_{luK}(-k_{y'}, \pi/6)$. Hence in general,
$T_{lu}$ and $T_{ld}$ at the same value of $\theta$ remain
distinct. However in the special case of unbiased BLG, Eq.~(\ref{symrel5})
together with the $2\pi/3$ periodicity that results from trigonal warping,
give $T(k_{y'}, \pi/6) = T(-k_{y'}, \pi/2)$. Then it follows that
$T_{ldK}(k_{y'}, \pi/6) = T_{luK}(k_{y'}, \pi/6)$ and
$T_{ldK}(k_{y'}, \pi/2) = T_{luK}(k_{y'}, \pi/2)$. Hence only one function
is needed and can be taken to be $T_{lu}$.

When there is no TW, the transmission coefficients are independent of
$\theta$ because the constant energy contours are circular. Then
reasoning similar to that used in the armchair case leads to
$T_{ldK}(k_{y'}) = T_{luK}(-k_{y'})$. Again only one function
is needed and can be taken to be $T_{lu}$.

In the exceptional angular range in Fig.~\ref{oblincfig} (left), one
incident state couples to two propagating transmitted states. When the step
is reversed this changes to two incident states each of which
couples to one propagating transmitted state. We have investigated this
case numerically for unbiased BLG as in Fig.~\ref{oblincfig} (left) at the
incidence conditions and potentials given in the figure caption. We find
that the sum of the transmission coefficients can be obtained from one
independent function and this function can be taken to be $T_{lu}$ as given
by Eq.~(\ref{tcoeffdef}) for the case when there are two propagating
transmitted states. We also find that the sum satisfies Eq.~(\ref{symrel2}).
When the sum is known in one valley, this equation gives the sum in the
other one. 

\subsection{Relations used in Section \ref{ValleySection}}
Eq.~(\ref{xreveq}) is a consequence of Eq.~(\ref{symrel3}) and the fact
that $T$ is independent of $\theta$ when there is no TW. Alternatively,
Eq.~(\ref{xreveq}) can be obtained from the $K$ Hamiltonian,
Eq.~(\ref{hblg}). When there is no TW, inverting the $x'$ co-ordinate,
i.e. putting $x'\rightarrow -x'$, transforms $H_K$ into the $K'$
Hamiltonian, $\hat{H}_{K'}$, in which the potentials $V_i(x')$ are replaced
by $V_i(-x')$. This leads to Eq.~(\ref{xreveq}).

\section{Pseudospin conditions for unbiased BLG}
\label{UnbiasedAppendix}

The pseudospin conservation condition can be stated in two ways. In the
first report of AK tunneling in BLG, \cite{Katsnelson06} the authors say
that the propagating states on the left side of a step match onto an
evanescent state on the right so both states have the same pseudospin. In
later reports, \cite{Gu11, Park11, Park12} the authors say equivalently
that the propagating states on opposite sides of the step are of opposite
pseudospin. These conditions result from swap symmetry and we show this by
using the approximate polarization vectors.

It is convenient to work in a representation where the
component order is non-dimer followed by dimer, i.e. the approximate
4-component states are of form $(\tilde{\phi}_{A1}, \tilde{\phi}_{B2},
\tilde{\phi}_{A2}, \tilde{\phi}_{B1})^T$. The approximate polarization
vectors for the evanescent (e) and propagating (p) states are
\begin{eqnarray}
  \tilde{\mathbf{e}}_e &=& N_e
  (1, \tilde{a}_e, -iv_0\hbar \tilde{\lambda}/t, -i\tilde{a}_e
  v_0\hbar\tilde{\lambda}/t)^T,\nonumber\\
  \tilde{\mathbf{e}}_p &=& N_p
  (1, \tilde{a}_p, -v_0\hbar \tilde{k}_x/t, -\tilde{a}_p v_0\hbar \tilde{k}_x/t)^T,
  \label{2cevecev}
\end{eqnarray}
where $i\tilde{\lambda}$ and $\tilde{k}_x$ are approximations to the
$x$-component of $\mathbf{k}$ and $N_i$ are normalization constants.
$\tilde{a}_i = \pm\mathrm{sgn}(E-V_1)\sqrt{(E-V_1)/(E-V_2)}$ where the sign is
$+$ for the evanescent state and $-$ for the propagating state.
The swap operator in the same representation is
\begin{equation}
  S = 
    \left( \begin{array}{cc}
    \sigma_x & 0\\
    0 & \sigma_x \\
    \end{array} \right). \label{swapop2c}
\end{equation}

In unbiased BLG, $\tilde{a}_e = \pm 1$, $\tilde{a}_p = \mp 1$, where the
upper signs apply in the conduction band and the lower signs apply in the
valence band.  Hence conduction band \textit{propagating} states are swap
antisymmetric ($s=-1$) and so are valence band \textit{evanescent
  states}. The 2-component vectors formed from these vectors by neglecting
the dimer components are eigenvectors of the pseudospin, $\sigma_x$, with
eigenvalue $s_x=s$. Thus the pseudospins on both sides of the step are
identical when the state on the right is purely evanescent. The pseudospin
condition on the propagating states can be obtained in a similar way.

\section{Pseudospin conditions for biased BLG}
\label{BiasedAppendix}

\subsection{Rotation of polarization vectors}
\label{RotationSection}

In biased BLG, the pseudospin condition for AK tunneling at a potential
step is that the incident state on the left side and the evanescent state
on the right side have the same the pseudospin expectation value. Or
equivalently, that the expectation values of the pseudospin of the
right propagating states on either side of the step are of equal magnitude
and opposite sign \cite{Park11}.

AK tunneling at normal incidence \cite{Park11} occurs when the potentials
and energy satisfy
\begin{eqnarray}
  \frac{E-V_{1l}}{E-V_{2l}} &=& \frac{E-V_{1r}}{E-V_{2r}},\nonumber\\
  \mathrm{sgn}(E-V_{1l}) &=& -\mathrm{sgn}(E-V_{1r}).
  \label{normcon2c}
\end{eqnarray}
The pseudospin expectation value conditions result from evaluating the
pseudospin expectation values for the 2-component states that occur when
Eq.~(\ref{normcon2c}) is satisfied.

To show these conditions and Eq.~(\ref{normcon2c}) result from swap
symmetry, we rotate the approximate 4-component polarization vectors for an
evanescent state so they become eigenstates of the swap operator. This
rotation can always be performed but we show that AK tunneling occurs only
for a critical pair of rotation angles. These angles give
Eqs.~(\ref{normcon2c}) and the pseudospin expectation condition.

The necessary rotation matrix is
\begin{equation}
  R = 
    \left( \begin{array}{cc}
    Q(\omega) & 0\\
    0 & Q(\omega) \\
    \end{array} \right), \label{rotop2c}
\end{equation}
where
\begin{eqnarray}
  Q &=&
    \left( \begin{array}{cc}
    \cos(\omega) & -\sin(\omega)\\
    \sin(\omega) & \cos(\omega) \\
    \end{array} \right),\nonumber\\
    \omega &=& \pm \frac{\pi}{4} - \tan^{-1} \tilde{a}_e.
    \label{rotopq}
\end{eqnarray}
Here the sign is that of the desired $S$ eigenvalue and the rotation angle
$\omega$ is chosen so that $\tilde{a}_e$ becomes $\pm 1$. Thus the rotated
vector becomes an eigenvector of $S$.

To identify the critical rotation angles it is convenient to work with only
the $\mathbf{e}$ vectors. We use
Eq.~(\ref{orthojdef}) to write Eq.~(\ref{approxantikcon}) as
\begin{equation}
  \tilde{\mathbf{e}}^\dagger_{4l} \tilde{v}_{x'} \tilde{\mathbf{e}}_{3r} = 0,
  \label{velcon2c}
\end{equation}
where the velocity operator in the (non-dimer, dimer) representation is
\begin{equation}
  \tilde{v}_{x'} = v_0 
    \left( \begin{array}{cc}
    0 & \sigma_x\\
    \sigma_x & 0 \\
    \end{array} \right). \label{velop2c}
\end{equation}

We choose the rotation angles $\omega_l$ and $\omega_r$ so that the $S$
eigenvalues on the left and right sides of the step are of opposite
sign. Then we insert these rotations into Eq.~(\ref{velcon2c}).  This gives
\begin{eqnarray}
  \tilde{\mathbf{e}}^\dagger_{4l} \tilde{v}_{x'} \tilde{\mathbf{e}}_{3r} &=&
  \tilde{\mathbf{e}}^\dagger_{4l} R^T(\omega_l) R(\omega_l) \tilde{v}_{x'}
  R^T(\omega_r) R(\omega_r)\tilde{\mathbf{e}}_{3r} \nonumber\\
  &=& \tilde{\mathbf{e}}^\dagger_{4l} R^T(\omega_l) R(\omega_l +\omega_r)
  \tilde{v}_{x'} R(\omega_r)\tilde{\mathbf{e}}_{3r},
  \label{rotcon2c}
\end{eqnarray}
where we have used $\tilde{v}_{x'} R^T = R \tilde{v}_{x'}$.

Next, we show that the right hand side of Eq.~(\ref{rotcon2c}) vanishes
when $\omega_l+\omega_r = 0$. We obtain
\begin{eqnarray}
  &&\tilde{\mathbf{e}}^\dagger_{4l} R^T(\omega_l) R(\omega_l +\omega_r)
  \tilde{v}_{x'} R(\omega_r)\tilde{\mathbf{e}}_{3r}\nonumber\\
  &=& \tilde{\mathbf{e}}^\dagger_{4l} R^T(\omega_l) R(\omega_l +\omega_r)
  S\tilde{v}_{x'}S R(\omega_r)\tilde{\mathbf{e}}_{3r}, \nonumber\\
  &=& \tilde{\mathbf{e}}^\dagger_{4l} R^T(\omega_l) S R^T(\omega_l +\omega_r)
  \tilde{v}_{x'}S R(\omega_r)\tilde{\mathbf{e}}_{3r}, \nonumber\\
  &=& -\tilde{\mathbf{e}}^\dagger_{4l} R^T(\omega_l)R^T(\omega_l +\omega_r)
  \tilde{v}_{x'}R(\omega_r)\tilde{\mathbf{e}}_{3r},
  \label{omegacon2c}
\end{eqnarray}
where we have used $RS = SR^T$ and the fact that the $S$ eigenvalues on
opposite sides of the step are of opposite sign.
$R(\omega_l +\omega_r) = R^T(\omega_l +\omega_r) = I$ when
$\omega_l+\omega_r = 0$ and then it follows from Eq.~(\ref{omegacon2c})
that the right hand side of Eq.~(\ref{rotcon2c}) vanishes.

Eq.~(\ref{rotopq}) shows that $\omega_l+\omega_r = 0$ when $\tilde{a}_{el}
= -\tilde{a}_{er}$ and this condition leads to Eq.~(\ref{normcon2c}) and
the associated condition on the sign of $E-V_1$. Further, when
$\tilde{a}_{el} = -\tilde{a}_{er}$, the expectation values of the swap
operator on the left and right sides of the step satisfy
$\tilde{\mathbf{e}}_{1l}^\dagger S\tilde{\mathbf{e}}_{1l} =
\tilde{\mathbf{e}}_{3r}^\dagger S\tilde{\mathbf{e}}_{3r}$ and these
expectation values are identical to the pseudospin expectation values.
The reason for the equality of the swap and pseudospin expectation values
is that non-dimer and dimer sub-vectors of the approximate 4-component
polarization vectors are proportional to each other.

Although we have used a rotation that makes the evanescent states
eigenstates of $S$, it is \textit{impossible} to find a rotation that makes
all the plane wave states eigenstates of $S$. The reason is that
transformation of the coefficient matrix in Eq.~(\ref{approxeigkx}) results
in a matrix (Section \ref{TransformSection}) that has one invariant
subspace of dimension 2 so only 2 of the 4 rotated states can be
eigenstates of $S$. A rotation similar to $Q$ is used in ref. \cite{Park11}
but appears to be applied only to the propagating states. The
transformation of the evanescent states, which requires a different
rotation angle, is not discussed and neither is the invariant subspace.

\subsection{Transformation of coefficient matrix}
\label{TransformSection}

The transformation of the coefficient matrix in Eq.~(\ref{approxeigkx}) and
the resulting invariant subspace are illustrated in this section with the
example of $s=+1$ evanescent states in the conduction band. Similar
subspaces occur in all other cases. We also show that it is impossible to
find a rotation that makes all the plane wave states eigenstates of $S$.

In biased BLG, the swap operator commutes with neither the Hamiltonian nor
the coefficient matrix. This means the swap operator and coefficient matrix
cannot share a complete set of eigenvectors. However, non-commuting
operators may share a subset of eigenvectors. This occurs in the present
case and results in the invariant subspace.

We perform 2 steps to demonstrate the existence of the invariant subspace
and show that it is 2-dimensional. First, we transform the coefficient
matrix with the rotation operator, $R$, in Eq.~(\ref{rotop2c}). Then we
express the transformed matrix in the basis formed by the eigenvectors of
the swap operator.

In the (non-dimer, dimer) representation the coefficient matrix in
Eq.~(\ref{approxeigkx}) becomes
\begin{equation}
  C = \frac{1}{v_0} 
    \left( \begin{array}{rrrr}
    0 & 0 & t & 0 \\
    0 & 0 & 0 & t \\
    0 & -\varepsilon_2 & 0 & 0 \\
    -\varepsilon_1 & 0 & 0 & 0 \\
    \end{array} \right), \label{coeffmatndd}
\end{equation}
where $\varepsilon_i = E - V_i$. This matrix is not swap symmetric because
$V_1 \ne V_2$ in biased BLG. The lack of swap symmetry persists after the
matrix has been transformed.

The matrix of eigenvectors of the swap operator in the (non-dimer, dimer)
representation is
\begin{equation}
  \frac{1}{\sqrt{2}} 
    \left( \begin{array}{rrrr}
    1 & 0 & 1 & 0 \\
    1 & 0 & -1 & 0 \\
    0 & 1 & 0 & 1 \\
    0 & 1 & 0 & -1 \\
    \end{array} \right), \label{swapbasis}
\end{equation}
where the order is two $s=1$ vectors followed by two $s=-1$ vectors.

The transformed matrix, expressed in the swap eigenvector basis, is
\begin{equation}
  C' = \frac{1}{v_0} 
    \left( \begin{array}{rrrr}
    0 & t & 0 & 0 \\
    -\alpha & 0 & \beta & 0 \\
    0 & 0 & 0 & t \\
    \gamma & 0 & \alpha & 0 \\
    \end{array} \right), \label{transmat}
\end{equation}
where
\begin{eqnarray}
  2 \alpha &=& (\varepsilon_1 + \varepsilon_2) \cos 2\omega,\nonumber\\
  2 \beta &=&  (\varepsilon_1 + \varepsilon_2) \sin 2\omega -
  (\varepsilon_1 - \varepsilon_2),\nonumber\\
  2 \gamma &=& (\varepsilon_1 + \varepsilon_2) \sin 2\omega +
  (\varepsilon_1 - \varepsilon_2).
  \label{abgdef}
\end{eqnarray}
Eqs.~(\ref{transmat}) and (\ref{abgdef}) are valid for arbitrary $\omega$.

We now show that the transformed matrix has an invariant subspace when
$\omega$ is chosen so that the evanescent wave polarization vectors are
rotated so they become eigenvectors of the swap operator. In the case of 
the $s=+1$ subspace in the conduction band, Eqs.~(\ref{rotopq}) and
(\ref{abgdef}) give
\begin{eqnarray}
  \alpha &=& \sqrt{\varepsilon_1 \varepsilon_2}, \nonumber\\
  \beta &=& \varepsilon_2 - \varepsilon_1,\nonumber\\
  \gamma &=& 0.
  \label{specabg}
\end{eqnarray}
As $\gamma = 0$, the lower left $2\times 2$ sub-matrix of the transformed
matrix vanishes, hence the space spanned by the $s=+1$ vectors forms an
invariant subspace of dimension 2, as stated in Section \ref{RotationSection}.

The eigenvectors that span this invariant subspace are of form
$(u_1, u_2, 0, 0)^T$ and satisfy
\begin{equation}
    \left( \begin{array}{rr}
    0 & t \\
    -\alpha & 0\\
    \end{array} \right)
    \left( \begin{array}{r}
    u_1\\
    u_2\\
    \end{array} \right) =    
    -iv_0 \hbar \tilde{\lambda}\left( \begin{array}{r}
    u_1\\
    u_2\\
    \end{array} \right).
    \label{invarvec}
\end{equation}
The eigenvalues are $\pm i\sqrt{t \sqrt{\varepsilon_1 \varepsilon_2}}$ and
give the known values of $\tilde{\lambda}$ in the 2-component
approximation. The remaining 2 eigenvectors of $M'$ are propagating states
with a mixture of $s=+1$ symmetry and $s=-1$ symmetry. Replacing
$\tilde{a}_e$ with $\tilde{a}_p = -\tilde{a}_e$ in Eq.~(\ref{rotopq}) gives a
transformation that puts the propagating states in the invariant subspace
and makes the evanescent states a mixture of symmetry types. Hence it is
impossible to find \textit{one} value of $\omega$ that makes \textit{all}
the states eigenstates of $S$, as stated in Section \ref{RotationSection}.

\section{Tight binding theory of AK tunneling}
\label{TBAppendix}

We use tight binding theory to find the transmission and reflection
coefficients for Bloch waves at normal incidence on an armchair step in
unbiased BLG. AK tunneling occurs in this situation because of swap
symmetry and the transmission and reflection coefficients are almost
identical to those found with the continuum theory.

\begin{figure}
  \begin{center}
  \vspace{5mm}
  \includegraphics[width=7.0cm]{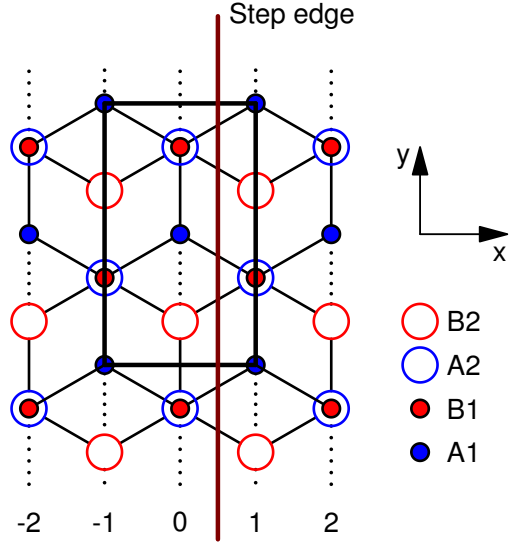}
  \caption{(Color online). Armchair step geometry. Bold rectangle: unit
    cell; labeled dotted lines: atomic columns; bold brown vertical line:
    step edge. The site labels are as in Section \ref{TheorySection}.}
\label{stepfig}
\end{center}
\end{figure}

Fig.~\ref{stepfig} shows the step geometry. We use a rectangular unit cell
that has twice the area of the primitive cell. The atoms are arranged in
columns separated by a distance $a/2$, where $a$ is the lattice
constant. There are 2 columns per cell and we take the cell origins to be
on the even-numbered columns. Each column contains 4 inequivalent
sites. The step edge is midway between columns 0 and 1. The midway position
ensures that potential does not change abruptly at any atomic site.

The tight binding Bloch waves are a superposition of basis Bloch waves:
\begin{equation}
  \phi_\mathbf{k} = \frac{1}{\sqrt{N}} \sum_s v_s \sum_{\mathbf{R}}
  e^{i \mathbf{k} \cdot (\mathbf{R} + \mathbf{d}_s)}
  u(\mathbf{r} - (\mathbf{R} + \mathbf{d}_s)),
  \label{Blochdef}
\end{equation}
where $N$ is the number of cells. The cell origins are at positions
$\mathbf{R}$, the position of site $s$ in the unit cell is
$\mathbf{d}_s$ and $u$ is an atomic orbital. The sum over $\mathbf{R}$ is a basis Bloch wave and the
numbers $v_s$ are expansion coefficients. These coefficients are the
elements of a polarization vector, $\mathbf{v}$.

Normal incidence on an armchair step corresponds to incidence in the
crystallographic $x$ direction (Fig. \ref{stepfig}). Hence the equation for
$\mathbf{v}$ can be obtained by putting $k_y = 0$ in
the $\mathbf{k}$-space Hamiltonian in ref. \cite{McCann13}. This gives
\begin{equation}
  \left\lbrace\left[A + (V' - E) I \right]  + \lambda_x A + \lambda_x^{-1}
  A\right\rbrace \mathbf{v} = \mathbf{0},
  \label{qep}
\end{equation}
where 
\begin{equation}
  A = \left(
  \begin{array}{rrrr}
    0 & -\gamma_0 & \gamma_4 & -\gamma_3\\
    -\gamma_0 & 0 & \gamma_1 & \gamma_4\\
    \gamma_4 & \gamma_1 & 0 & -\gamma_0\\
    -\gamma_3 & \gamma_4 & -\gamma_0 & 0
  \end{array}
  \right),
  \label{amatdef}
\end{equation}
is the matrix of tight binding parameters and $I$ is the $4\times 4$ unit
matrix. $V' = \mathrm{diag}(V, V + \Delta', V + \Delta', V)$, where $V$ is
the potential. The Hamiltonian for $k_y = 0$ is swap symmetric because $A$
and $V'$ are swap symmetric.

At fixed energy, Eq.~(\ref{qep}) represents a quadratic eigenvalue problem
(QEP) for $\lambda_x \equiv \exp(ik_x a / 2)$. This is not the only way of
finding $\lambda_x$ as one can instead \cite{Chen16} write Eq.~(\ref{qep})
as a linear eigenvalue problem for $(\lambda_x + \lambda_x^{-1})$. However
the QEP formulation is better for our purposes because it leads directly to
an orthogonality condition analogous to Eq.~(\ref{antikcon}).

The QEP defined by Eq.~(\ref{qep}) is palindromic \cite{Mackey06} and this
property guarantees that the plane waves occur in $\pm k_x$ pairs. QEPs can
normally be solved numerically with a linearization method, however a
special linearization is needed to preserve the $\pm k_x$ pairing. We
use the linearization recommended in ref. \cite{Mackey06} and write our QEP
as
\begin{equation}
  \left[
  \left(
  \begin{array}{cc}
    A & A\\
    A_0 - A & A
  \end{array}
  \right)
  + \lambda_x
  \left(
  \begin{array}{cc}
    A & A_0 - A\\
    A & A
  \end{array}
  \right)
  \right]
  \left(
  \begin{array}{c}
    \lambda_x \mathbf{v}\\
    \mathbf{v}
  \end{array}
  \right) = \mathbf{0},
  \label{linqep}
\end{equation}
where $A_0 = A + (V' - E) I$.

The solution of the non-symmetric eigenvalue problem (\ref{linqep}) gives 8
right polarization vectors, $\mathbf{e}$, of form $\mathbf{e}^T = (\lambda_x
\mathbf{v}^T, \mathbf{v}^T)$. Because of the $\pm k_x$ pairing, 4 of these
vectors are associated with the $K$ valley and 4 with the $K'$ valley. The
physical meaning of the $\mathbf{e}$ vectors is that the first 4 components are Bloch
wave amplitudes on column 1 and the last 4 are the amplitudes on column 0.
As the eigenvalue problem is nonsymmetric, the solution also gives a set of
left polarization vectors, $\mathbf{f}^\dagger$. The $\mathbf{e}$ and
$\mathbf{f}^\dagger$ vectors form a biorthogonal set as described in
Section \ref{4cHPWSection}.

The wave functions on the left and right sides of the step are
\begin{eqnarray}
  \psi_l &=& 
    \phi_{k_{1l\tau_i}} +
    \sum_{\tau} r_{2\tau} \phi_{k_{2l\tau}} + r_{4\tau} \phi_{k_{4l\tau}},\\
  \label{tbpsileft}
  \psi_r &=&
    \sum_{\tau} t_{1\tau} \phi_{k_{1r\tau}} + t_{3\tau} \phi_{k_{3r\tau}},
  \label{tbpsiright}
\end{eqnarray}
where the notation is similar to that in Eqs.~(\ref{psileft}) and
(\ref{psiright}). However Bloch waves replace the plane waves and $\psi_l$
and $\psi_r$ are formed from Bloch waves from both valleys to account for
the possibility of valley mixing. $\tau$ is the valley index and $\tau_i$
is the valley of incidence.  The system wave function is
$\psi = \psi_l$ when $x < a/4$ and $\psi = \psi_r$ when $x > a/4$.

Equations for the transmission and reflection coefficients are obtained
from the condition \cite{Osbourn79} that $\psi$ is an eigenstate of the
tight binding Hamiltonian, $H_{TB}$, that is $(H_{TB} - E)|\psi\rangle = 0$.
This condition is satisfied when
\begin{equation}
  \langle u(\mathbf{R}_s)|(H_{TB} - E)|\psi\rangle = 0,
  \label{tbcond1}
\end{equation}
for each of the 8 atomic sites, $\mathbf{R}_s$, adjacent to the step
edge. No other sites need to be considered as the in-plane coupling is
restricted to nearest neighbors. Eqs.~(\ref{tbcond1}) give 8 linear
equations for the 4 unknown transmission coefficients and the 4 unknown
reflection coefficients.

Eqs.~(\ref{tbcond1}) are linear in the amplitudes of $\psi_l$ and $\psi_r$
at the site $\mathbf{R}_s$. The site amplitude of a Bloch wave at site $s$
in column $n$ is $v_s \exp(i k_x na / 2)$, as can be seen from
Eq.~(\ref{Blochdef}). After some tedious manipulations involving these site
amplitudes, it can be shown that Eqs.~(\ref{tbcond1}) are equivalent to the
simpler condition that the site amplitudes in $\psi_l$ and $\psi_r$ are equal on
column 0 and equal on column 1 \cite{sitenote}. This condition can be written
as the vector equation
\begin{equation}
  \mathbf{e}_{1l\tau_i} +
  \sum_{\tau} r_{2\tau} \mathbf{e}_{2l\tau} + r_{4\tau} \mathbf{e}_{4l\tau} =
  \sum_{\tau} t_{1\tau} \mathbf{e}_{1r\tau} + t_{3\tau} \mathbf{e}_{3r\tau},
  \label{tbpsimatch}
\end{equation}
where the vectors $\mathbf{e}$ are the 8-component polarization vectors
found by solving Eq.~(\ref{linqep}). Eq.~(\ref{tbpsimatch}) is the tight
binding analog of Eq.~(\ref{psimatch}). We solve it with the
biorthogonality method we used to solve Eq.~(\ref{psimatch}).

By following the same steps that led to Eq.~(\ref{antikcon}), we find that
$t_{1\tau}$ vanishes in both valleys when
\begin{equation}
  \mathbf{f}^\dagger_{3lK}\cdot\mathbf{e}_{3rK} =
  \mathbf{f}^\dagger_{3lK'}\cdot\mathbf{e}_{3rK'} =
  \mathbf{f}^\dagger_{3lK}\cdot\mathbf{e}_{3rK'} =
  \mathbf{f}^\dagger_{3lK'}\cdot\mathbf{e}_{3rK} =
  0.\label{tbantikcon}
\end{equation}
These scalar products vanish because of swap symmetry as in the continuum
approach. The swap eigenvalues of the Bloch states are identical in both
valleys because the matrix $A$ in Eq.~(\ref{amatdef}) is
$\mathbf{k}$-independent. Hence the swap classification of the propagating
and evanescent Bloch waves is the same as the plane wave swap
classification found in Section \ref{SymmetrySection}. The 8-component
$\mathbf{f}^\dagger$ and $\mathbf{e}$ vectors have the same swap
eigenvalues as the Bloch waves because the matrices in Eq.~(\ref{linqep})
are invariant under the 8-component swap operator
\begin{equation}
  S_8 =
  \left(
  \begin{array}{cc}
    S & 0\\
    0 & S
  \end{array}
  \right).
\end{equation}
$S_8^2 = I_8$, the $8 \times 8$ unit matrix. Hence for any pair of 
$\mathbf{f}^\dagger$ and $\mathbf{e}$ vectors with opposite swap
eigenvalues, $\mathbf{f}^\dagger \cdot \mathbf{e} =
\mathbf{f}^\dagger S_8^2 \cdot \mathbf{e} =
-\mathbf{f}^\dagger \cdot \mathbf{e}$. Therefore all the scalar products in
Eq.~(\ref{tbantikcon}) vanish in the energy range where the incident
state is in the conduction band and the transmitted state is in the
valence band or vice versa. Thus AK tunneling occurs in the same energy
range as found in the continuum approach (Section \ref{SymmetrySection}).

\begin{figure}[H]
  \vspace{3mm}
  \begin{center}
    \hspace{-4mm}
  \includegraphics[width=4.1cm, angle=-90]{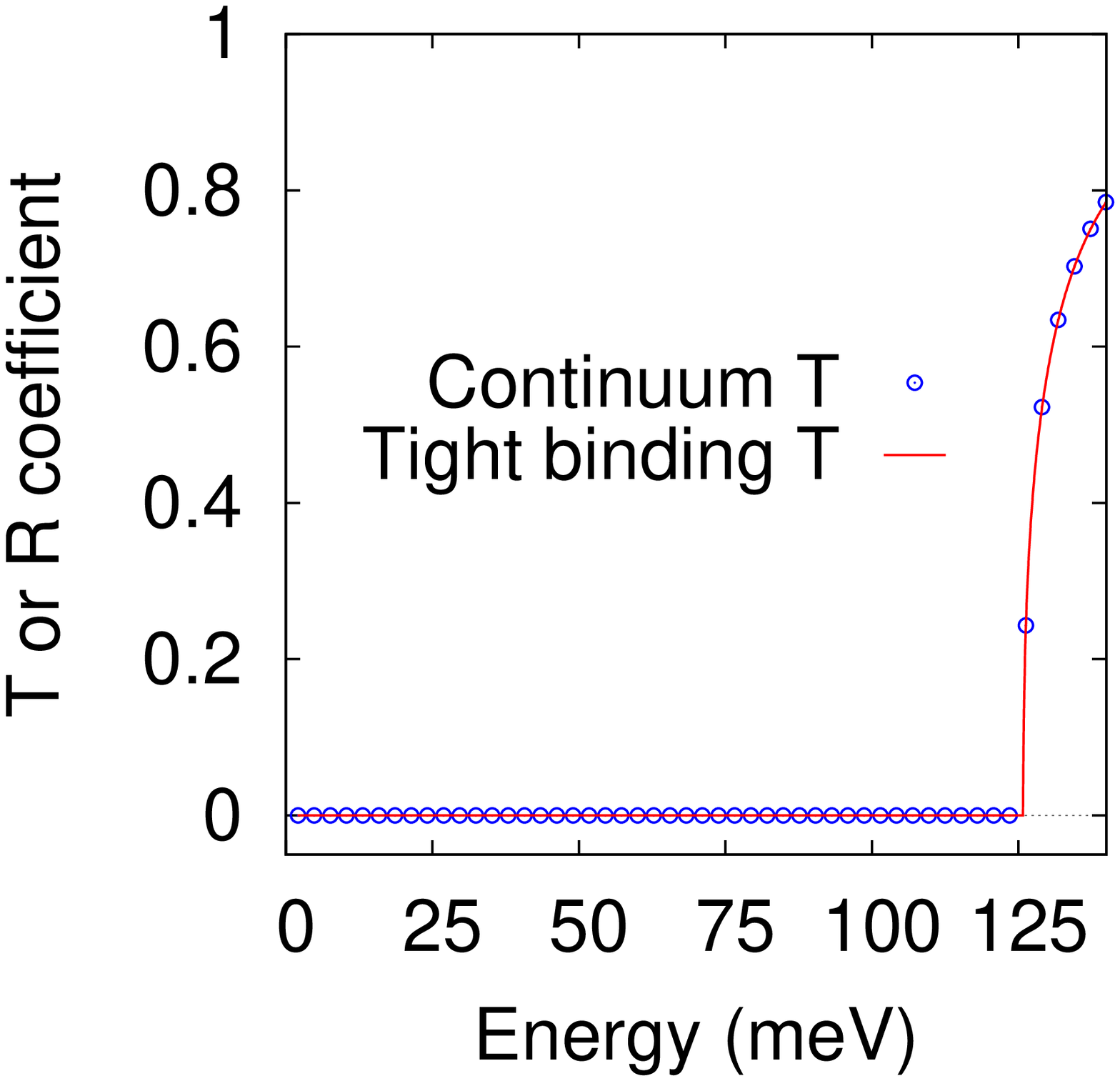}
    \hspace{3mm}
  \includegraphics[width=4.1cm, angle=-90]{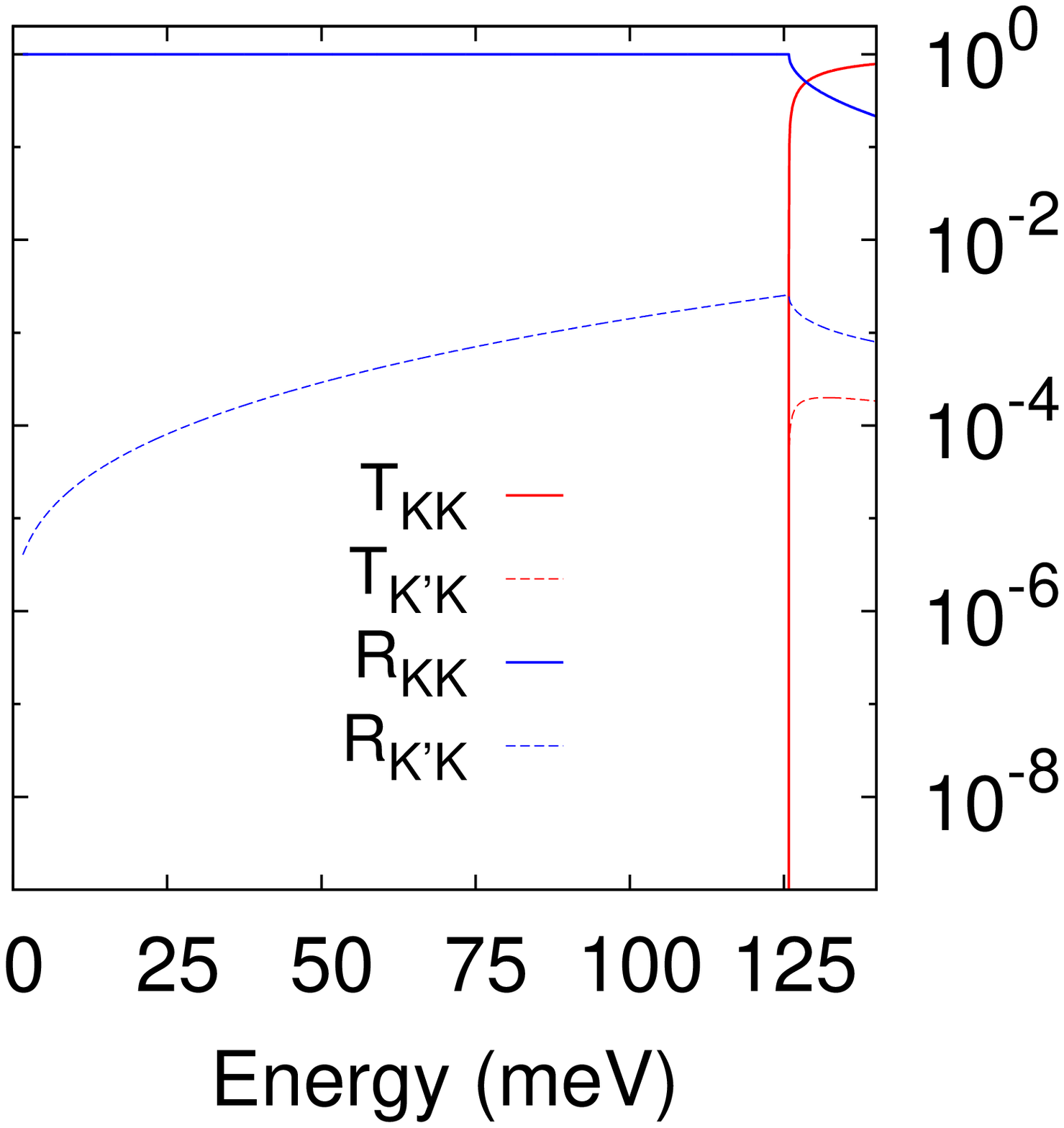}
  \caption{(Color online). Tight binding transmission ($T$) and
    reflection ($R$) coefficients for normal current incidence in the $K$
    valley on an armchair step of height 127 meV in unbiased BLG.
    Left: comparison of tight binding and continuum $T$. In the tight
    binding case $T = T_{KK} + T_{K'K}$. The first subscript is the
    output valley and the second subscript is the input valley.
    Right: valley resolved tight binding $T$ and $R$. }
\label{tbnormincfig}
\end{center}
\end{figure}

Fig.~\ref{tbnormincfig} (left) shows the excellent agreement between
transmission coefficients computed with the continuum and tight binding
approaches. The difference between the transmission coefficients is at most
$\simeq 6\times 10^{-4}$ at $E \simeq 135$ meV. Fig.~\ref{tbnormincfig}
(right) shows that the valley mixing is very small. The valley-flip
transmission and reflection coefficients are typically between 3 and 5
orders of magnitude smaller than the valley-preserving coefficients. 
Similar small valley mixing was reported in earlier work on barrier
transmission away from the anti-Klein condition \cite{Chen16}.


\begin{thebibliography}{99}
\bibitem{footnote1} One exception is the case of normal incidence on an
  armchair step in unbiased BLG.
\bibitem{sitenote} Eqs.~(\ref{tbcond1}) can be rearranged into a homogeneous
  system of linear equations for the differences of the site amplitudes in
  $\psi_l$ and $\psi_r$. These equations have only the trivial solution
  unless $\gamma_0 = \gamma_4$, which is not the case.
\end{thebibliography}
\end{document}